\documentclass[letterpaper,11pt]{article}
\usepackage[margin=2.3cm]{geometry}

\usepackage{tgtermes}

\usepackage[utf8]{inputenc}
\usepackage[T1]{fontenc}

\usepackage{graphicx}
\usepackage[table]{xcolor}
\usepackage{enumerate}
\usepackage[inline]{enumitem}

\usepackage{bm}
\usepackage{amsmath}
\usepackage{amssymb}

\usepackage[title,toc]{appendix}
\usepackage{booktabs}
\usepackage[perpage]{footmisc} 
\usepackage{multicol}
\usepackage{multirow}
\usepackage{float}
\usepackage{array}
\newcolumntype{C}{>{$}c<{$}}

\usepackage{indentfirst}
\usepackage{parskip}
\usepackage[
labelfont={small,bf},
singlelinecheck=true,
textfont={small},
justification=justified
]{caption}

\usepackage[justification=raggedright,
singlelinecheck=false]{subcaption}
\usepackage[
bibencoding=utf8,
giveninits=true,
sorting=none,
url=false,
doi=false,
maxbibnames=10,
backend=biber,
]{biblatex}

\AtBeginBibliography{}

\usepackage[hidelinks,
colorlinks=true,
citecolor={blue},
linkcolor={blue},
urlcolor={blue}]{hyperref}
\hypersetup{pdfpagemode=UseNone} 
\usepackage[noabbrev,capitalize]{cleveref}
\usepackage{titlesec}
\titleformat*{\section}{\bfseries\Large}
\titleformat*{\subsection}{\bfseries}
\titleformat*{\subsubsection}{\bfseries}
\titleformat*{\paragraph}{\bfseries}
\titleformat*{\subparagraph}{\bfseries}
\graphicspath{{./images/}}
\makeatletter
\DeclareRobustCommand*{\bfseries}{%
	\not@math@alphabet\bfseries\mathbf
	\fontseries\bfdefault\selectfont
	\boldmath
}
\makeatother
\DeclareSymbolFont{myletters}{OML}{ztmcm}{m}{it}
\DeclareMathSymbol{\lambda}{\mathord}{myletters}{"15}

\newcommand{\DPP}[2]{\frac{\partial{#1}}{\partial{#2}}}

\newcommand{\grad}[1]{\bm{\nabla}{#1}}

\newcommand{\br}{\mathbf{r}}

\newcommand{\be}{\bm{e}}

\newcommand{\bF}{\mathbf{F}}

\newcommand{\bT}{\mathbf{T}}
\newcommand{\bu}{\mathbf{u}}

\def\Re{Re}

\def\Rsq{R^2}

\renewcommand{\leq}{\leqslant}

\begin{document}
\title{\Large\bfseries{Physics-inspired architecture for neural network modeling of forces and torques in particle-laden flows}}

\author{{Arman Seyed-Ahmadi}\textsuperscript{1,2}
\& {Anthony Wachs}\textsuperscript{2,3}}
\date{}
\maketitle
\begin{center}
\begin{small}
\parbox{0.8\textwidth}{\centering\itshape\footnotesize
\textsuperscript{1}{Department of Statistics, University of British Columbia,\\ 2207 Main Mall, Vancouver, BC, V6T 1Z4, Canada}\\
\textsuperscript{2}{Department of Chemical
\& Biological Engineering, University of British Columbia,\\ 2360 East Mall, Vancouver, BC, V6T 1Z3, Canada}\\
\textsuperscript{2}{Department of Mathematics, University of British Columbia,\\ 1984 Mathematics Road, Vancouver, BC, V6T 1Z2,
 Canada}}
\end{small}
\end{center}

\begin{abstract}
\noindent We present a physics-inspired neural network (PINN) model for direct prediction of hydrodynamic forces and torques experienced by individual particles in stationary beds of randomly distributed spheres. In line with our findings, it has recently been demonstrated that conventional fully connected neural networks (FCNN) are incapable of making accurate predictions of force variations in a static bed of spheres \cite{Balachandar2020}. The problem arises due to the large number of input variables (i.e., the locations of individual neighboring particles) leading to an overwhelmingly large number of training parameters in a fully connected architecture. Given the typically limited size of training datasets that can be generated by particle-resolved simulations, the NN becomes prone to missing true patterns in the data, ultimately leading to overfitting. Inspired by our observations in developing the microstructure-informed probability-driven point-particle (MPP) model \cite{SeyedAhmadi2020}, we incorporate two main features in the architecture of the present PINN model: 1) superposition of pairwise hydrodynamic interactions between particles, and 2) sharing training parameters between NN blocks that model neighbor influences. These strategies helps to substantially reduce the number of free parameters and thereby control the model complexity without compromising accuracy. We demonstrate that direct force and torque prediction using NNs is indeed possible, provided that the model structure corresponds to the underlying physics of the problem. For a Reynolds number range of $ 2 \leq \Re \leq 150 $ and solid volume fractions of $ 0.1 \leq \phi \leq 0.4 $, the PINN's predictions prove to be as accurate as those of other microstructure-informed models.
\end{abstract}
\section{Introduction}\label{sec:introduction}
The astonishingly rapid advancement of computer hardware and growth of computational power over the past two decades has entirely transformed science and engineering. In fluid dynamics research, high-performance computing today plays an indispensable role in both the generation and processing of data. While direct numerical simulation tools provide nearly assumption-free models of fluid flow by directly solving the discretized Navier-Stokes (NS) equations, the cost of resolving all spatio-temporal scales becomes prohibitively high when multiscale phenomena are present. An important case in point is the numerical modeling of particle-laden flows, which is notoriously challenging for realistic system sizes due to their well-known multiscale nature, combined with a lack of scale separation. Particle-resolved direct numerical simulations (PR-DNS) are able to generate physically accurate results via direct computation of particle-fluid interaction forces and torques from high resolution field data. This is practically impossible for many real-world applications of particulate flow systems which typically contain billions of particles.
A viable alternative to PR-DNS is to average NS equations over length scales larger than the particle diameter $ d $, while still keeping track of particles in a Lagrangian manner \cite{Anderson_1969,Capecelatro2013}. This approach---commonly referred to as the Euler-Lagrange (EL) or point-particle technique---avoids resolving particle interfaces, albeit at the expense of the need for closure modeling of particle-fluid interactions. Additionally, sub-grid stresses induced by velocity fluctuations are also filtered as a result of the averaging process, which means that their modeling is necessary if physical fidelity is to be maximally retained \cite{Capecelatro2013}. It is thus clear that the reliability of an EL simulation is critically dependent on the accuracy of the employed closure models for both interphase momentum exchange and the filtered fluid fluctuations.

For a dilute suspension of particles (i.e. $ \phi \to 0 $, with $ \phi $ being the solid volume fraction) in the Stokes limit (i.e. $ \Re \to 0 $, where $ \Re $ shows the Reynolds number), the Maxey-Riley-Gatignol (MRG) equation \cite{Maxey1983,gatignol1983faxen} yields accurate predictions of the hydrodynamic forces \cite{Balachandar2010}. This equation generalizes the Faxén's law for spatially and temporally varying undisturbed flows, accounting for the quasi-steady, stress divergence, added-mass and viscous history forces \cite{Subramaniam2018}. While the MRG equation is rigorously valid for $ \Re = 0 $, its applicability can be extended to higher Reynolds numbers by introducing empirical corrections. The standard drag law of a sphere given as $ C_D = (24 / \Re) (1 + 0.15 \Re^{0.687}) $ \cite{schiller1933grundlegenden} is one such correction for the quasi-steady drag in the MRG equation for inertial regimes. In addition, shear- and rotation-induced lift forces---i.e. the Saffman \cite{saffman_1965} and Magnus \cite{Seifert2012} forces, respectively---are nonlinear force contributions arising in finite $ \Re $ regimes which also need to be modeled.
When a suspension of particles can no longer be regarded as dilute, the disturbance created by a particle is likely to influence the hydrodynamics of another particle due to the relatively small inter-particle distances. As a result, the flow around particles varies on the scale of the particle diameter, i.e. $ d \approx \eta $ where $ \eta $ denotes the length scale of the carrier flow variations. In such a system, the application of the MRG equation becomes problematic. This is because the force contributions are expressed in terms of the undisturbed flow, the computation of which is far from trivial in the presence of nearby particles. Moreover, the utilization of the Faxén's law in the MRG equation is justified under the assumptions of $ d \ll \eta $ and $ \Re \to 0 $ \cite{Maxey1983}, both of which are violated in an inertial suspension of finite-size particles.

The foregoing challenges with the finite-size point-particle approach has led to the development of drag models that account for the presence of other particles via introducing functional dependence on the solid volume fraction. While theoretical \cite{Batchelor1972,SANGANI1982343} and experimental \cite{Wen1966a,Felice1994} models with limited applicability in terms of $ \Re $ and $ \phi $ ranges have been available for decades, several drag correlations have been proposed more recently based on PR-DNS of stationary arrays of randomly dispersed spheres \cite{Beetstra2007,Tenneti2011,Tang2014,Bogner2015}. Despite their significant practical utility, correlations of this type are, at best, capable of providing only an average measure of the drag force that is collectively experienced by the particles. It is now well established that the particle-to-particle variation of the hydrodynamic forces due to the unique neighborhood of each particle is quite substantial \cite{Akiki_2016}, often being on the scale of the mean drag itself. Furthermore, the lack of a microstructure-dependent contribution precludes the computation of neighborhood-induced lift and torque altogether. These shortcomings appear to be major missing pieces of the puzzle when physical fidelity of EL simulations is concerned. It has been shown that conventional drag models employed in fluidized bed simulations significantly underestimate both the mean drag \cite{Kriebitzsch2013} and its variance \cite{Esteghamatian2017,Esteghamatian2017a}, the former being reflected in lower bed height values and the latter in smaller particle velocity fluctuations compared to PR-DNS.

The first major effort to deterministically account for the neighborhood effects on the hydrodynamic forces and torques of individual particles in a stationary dense array was the pairwise interaction extended point-particle (PIEP) model put forth by Akiki et al. \cite{Akiki_2017,Akiki2017}. The PIEP model first approximates the undisturbed flow by linearly superposing the perturbations created by each neighboring particle. The resulting non-uniform undisturbed flow is then used to obtain various force and torque contributions via the Faxén form of each term. Akiki et al. \cite{Akiki_2017,Akiki2017} showed that the PIEP model is able to predict $ \approx 40\% - 75\% $ of the force and torque variations for $ 0.11 \leqslant \phi \leqslant 45 $ and $ 40 \leqslant \Re \leqslant 173 $. The PIEP model is a purely physics-based approach that is constructed exclusively upon physical arguments and first principles of fluid mechanics.
Alternatively, the quest for a microstructure-dependent force/torque model may also be viewed as the problem of mapping a set of inputs $ \bm{x} \in \mathbb{R}^{d \times N} $ (e.g. neighbor locations) to a related set of outputs $ \bm{y} \in \mathbb{R}^{r \times N} $ (e.g. force/torques experienced a particle) via an appropriate function $ \bm{\mathcal{F}} $ of unknown parameters $ \bm{w} $ such that $ \bm{y} = \bm{\mathcal{F}}( \bm{x}; \bm{w} ) $. Note that $ d $ and $ r $ denote the dimensions of the input and output spaces, respectively, whereas $ N $ shows the number of samples in the dataset. Having specified a particular structure for $ \bm{\mathcal{F}} $, a data-driven approach typically seeks to fine-tune the parameters $ \bm{w} $ in a way that the overall error between true values and those predicted by the model is minimized.
An example of the data-driven strategy is the model of Moore et al. \cite{Moore2019} which employs a particular functional form for regression expanded in terms of spherical harmonics, the parameters of which are found through a non-linear optimization algorithm. They ultimately complement their data-driven model with the original physics-driven PIEP approach for performance improvements.
In our own effort to develop a data-driven neighborhood-dependent closure model, we proposed the microstructure-informed probability-driven point-particle (MPP) approach \cite{SeyedAhmadi2020}. Using PR-DNS data of stationary arrays of particles, the MPP model first identifies consistent, non-random patterns of neighboring particle locations according to a selective data filtering strategy. Based on statistical arguments, these patterns are used to infer physical basis functions to correlate neighbor locations with hydrodynamic forces and torques. The MPP model was found to have a performance similar to that of the PIEP hybrid model of \cite{Moore2019}. This similarity should not be surprising, given that both models---although in different ways---take advantage of the important approximation that the influence of individual neighbors on the force or torque of a test particle can be superposed. Despite being data-driven, both PIEP and MPP models are carefully formulated to preserve and incorporate physical features of the problem. In fact, most of the modeling effort of these approaches is expended on the structuring of the model, rather than the procedure for finding unknown parameters. Recently, Balachandar et al. \cite{Balachandar2020} proposed the concept of \textit{superposable wakes}. In this approach, they seek perturbation flows around spheres such that are assumed to be superposable. The superposition of these unknown perturbation flows are then used to compute forces and torques by utilizing the generalized Faxén's law. Finally, the perturbation flows are found by minimization of the error between the predicted forces and torques and the exact values from PR-DNS. Similar to \cite{Moore2019}, the form of the unknown functions---perturbation flows in this case---was assumed to be axisymmetric and hence expandable in spherical harmonics. Interestingly, the superposable wake model also shows the same performance characteristics as that of the PIEP and MPP models.

An area of active exploration in computational physics of fluids has been the application of machine learning (ML) techniques to tackle challenging existing problems. ML algorithms are an important category of data-driven approaches that offer great versatility for the mapping of complex nonlinear relationships without making strong prior assumptions about the model structure, i.e. the functional forms of the relationships between inputs and outputs. ML algorithms achieve this level of generality at the expense of their high demand for data and computational power required for their training. Among the most successful ML algorithms are neural networks (NN) which have been vastly applied to flow physics problems in turbulence modeling \cite{Xie2020,Yang2019,Beck2019,Xie2019,Fukami2019,Wu2018}, while less extensively so to direct flow prediction \cite{Lee2019,Sekar2019,Srinivasan2019,Leer2020,Raissi2018}, and multiphase flows \cite{Balachandar2020,He2019,Jiang2019,Ma2015,Ma2016} as well. The interested reader is referred to \cite{Brunton2020,Duraisamy2019,Brenner2019} for detailed reviews on the state of the art and perspectives of machine learning applications in fluid mechanics. The present work is inspired by our observation, consistent with the recent findings of Balachandar et al. \cite{Balachandar2020}, that a conventional fully connected NN cannot be successfully trained for direct prediction of drag variation based on the local neighborhood around each particle. The problem arises due to the relatively small size of datasets generated from PR-DNS containing a few thousand data points at best. For the highly dimensional problem of force/torque prediction based on local microstructure, such datasets are prohibitively underpopulated to be used by conventional NNs.
To remedy the existing issues, we devise a novel physics-inspired NN point-particle model that utilizes a custom architecture based on the problem's physical formulation to effectively guide the training procedure and overcome the high dimensionality issue. To this end, we incorporate the superposition approximation in the structure of the NN and share the trainable parameters between the NN blocks, each of which is responsible for taking into account the influence of a single neighboring particle.
In the remainder of this paper, we will first present the configuration of PR-DNS cases, followed by description of the model formulation and utilized assumptions and constraints. Subsequently, the structure of the proposed NN architecture is discussed in detail and is contrasted with conventional multilayer NNs. Finally, the performance of the model is evaluated and compared with multilayer NNs, and different aspects of model interpretability are explored.
\section{PR-DNS and data preparation}
In this work, we consider stationary random arrays of spheres dispersed in an incompressible Newtonian fluid, the flow of which is governed by the Navier-Stokes equations given below:
\begin{align}
	\DPP{\bu}{t} + \bu \cdot \grad{\bu} &= -\grad{p} + \frac{1}{\Re} \nabla^2 \bu \label{NS_eq},\\
	\grad \cdot \bu &= 0 \label{continuity},
\end{align}
where $ \bu $ and $ p $ show the fluid velocity vector and pressure. The Reynolds number denoted by $ \Re $ is defined as
\begin{align}\label{Re}
	\Re = \frac{\rho U d}{\mu} = \frac{\rho (1 - \phi) u_s d}{\mu},
\end{align}
with $ U $ and $ u_s $ representing the superficial and average interstitial velocities, whereas $ \rho $, $ \mu $ show the carrier fluid density and dynamic viscosity. Physical variables are non-dimensionalized using the particle diameter $ d $ as the length scale, $ U $ as the velocity scale and $ \rho U^2 $ as the pressure scale, $ \rho U^2 d^2 $ as the force scale and $ \rho U^2 d^3 $ as the torque scale.
Detailed descriptions of the numerical method employed to solve \cref{NS_eq,continuity} along with validation procedures are provided in \cite{SeyedAhmadi2020}. Each simulation is carried out in a tri-periodic cubic computational domain of edge length $ L $, inside of which $ N_p $ spheres are randomly distributed. The domain size $ L $ and the number of particles $ N_p $ are adjusted to achieve the desired solid volume fraction given as $ \phi $, while ensuring that each simulation generates an adequate amount of data. Also, a constant flow rate is imposed in the $ x $ direction to attain the target $ \Re $ via a dynamically adjusted pressure drop in our numerical algorithm.

\begin{table}[t]
	\centering
	\setlength{\tabcolsep}{15pt}
	\begin{tabular}[c]{C C C C C}
		\toprule
		\phi &	\Re	&	d / \Delta x	&	L  & N_p \\
		\midrule
		0.1 	&	2	&	24		& 	25		&	2984\\
		0.1 	&	10	&	24		& 	25		&	2984\\
		0.1 	&	40	&	24		& 	25		&	2984\\
		0.1 	&	150	&	32		& 	25		&	2984\\[6pt]
		0.2 	&	2	&	24		& 	20	 	&	3055\\
		0.2 	&	40	&	32		& 	20	 	&	3055\\
		0.2 	&	150	&	40		& 	20	 	&	3055\\[6pt]
		0.4 	&	2	&	32		& 	15		&	2578\\
		0.4 	&	40	&	40		& 	15		&	2578\\
		0.4 	&	150	&	48		& 	15		&	2578\\
		\bottomrule
	\end{tabular}
	\caption{Configurations of PR-DNS cases (adopted from \cite{SeyedAhmadi2020}) that are used for data generation in the present study. In the above table, $ \Delta x $ shows the grid spacing used by the numerical algorithm to discretize the governing equations \cref{NS_eq,continuity}.}
	\label{table:cases}
\end{table}
The PR-DNS cases that we use in the present study are identical to those in \cite{SeyedAhmadi2020}. The configuration of the respective cases are summarized in \cref{table:cases}.
The dataset for each case of \cref{table:cases} is constructed by identifying and recording the relative position vectors of the $ 30 $ closest neighbors of each particle in the array. These position vectors, together with the average local fluid velocity vector constitute the input variables (i.e. predictors) to our model, whereas the hydrodynamic force and torque components acting on each particle are the output variables.
\section{Model formulation}\label{sec:model_formulation}
We begin by defining a set of unit vectors for the subsequent development and analysis of the model. In what follows, $ \hat{\be}_x $, $ \hat{\be}_y $ and $ \hat{\be}_z $ show unit vectors along the three coordinate directions. Furthermore, $ \hat{\be}_\shortparallel $ denotes the direction of the pressure gradient driving the macroscale flow (i.e. the streamwise direction), whereas $ \hat{\be}_{\bot} $ represents any unit vector normal to $ \hat{\be}_\shortparallel $. In the present analysis, we take $ \hat{\be}_x $ to be coincident with $ \hat{\be}_\shortparallel $.
The force and torque on the particle $ i $ can be written as
\begin{subequations}
	\label{eq:decompose}
	\begin{align}
		\bF_i &= \langle \bF_i \rangle (\Re, \phi) + \Delta \bF_i(\Re, \phi, \{ \br_{j=1}, \dots, \br_{j=M} \}), \label{eq:F_decompose} \\
		\bT_i &= \Delta \bT_i(\Re, \phi, \{ \br_{j=1}, \dots, \br_{j=M} \}), \label{eq:T_decompose}
	\end{align}
\end{subequations}
where the angular brackets represent averaging over all particles in the array, while $ \Delta \bF_i $ and $ \Delta \bT_i $ show the deviations of the force and torque of a particle $ i $ from average values, respectively. Also, $ \{ \br_{j=1}, \dots, \br_{j=M} \} $ symbolizes the collection of position vectors of $ M $ influential neighbors relative to the position of particle $ i $. For each position vector $ \br_j $, we also define $ \hat{\be}_j $ to be the unit vector along the direction of $ \br_j $. The numbering $ j $ of neighboring particles is done according to their distance from particle $ i $, with $ j = 1 $ being the closest.

In a sufficiently large array, the lateral components of $ \langle \bF_i \rangle $ and all components of $ \langle \bT_i \rangle $ will tend to zero. However, particle-to-particle variation of forces and torques, i.e. $ \Delta \bF_i $ and $ \Delta \bT_i $, will be substantial. Conventional drag models provide empirical correlations for the variation of $ \langle \bF_i \rangle \cdot \hat{\be}_\shortparallel $ as a function of $ \Re $ and $ \phi $ \cite{Beetstra2007,Tenneti2011,Tang2014,Bogner2015}. Similar to our previous effort in developing a microstructure-informed force and torque model \cite{SeyedAhmadi2020}, here we focus on the deviations of the hydrodynamic forces and torques from average values. That is, we seek to approximate the functional dependence of $ \Delta \bF_i $ and $ \Delta \bT_i $ on $ \{ \br_{j=1}, \dots, \br_{j=M} \} $, the local microstructure of each particle.
In its most general form, the deviation terms $ \Delta \bF_i $ and $ \Delta \bT_i $ for particle $ i $ can be expressed as the sum of influences of all other $ N_p - 1 $ particles within the array, which may be written as
\begin{subequations}
	\label{eq:general_form}
	\begin{align}
		\Delta \bF_i &= \sum_{j=1}^{N_p - 1} \Delta \bF_{j \to i},\\
		\Delta \bT_i &= \sum_{j=1}^{N_p - 1} \Delta \bT{j \to i},
	\end{align}
\end{subequations}
recognizing that each influence term is a unique function of the relative positions of itself and all other neighbors, in addition to the flow regime parameters:
\begin{subequations}
	\label{eq:general_form2}
	\begin{align}
		\Delta \bF_{j \to i} &= \bm{f}_{j \to i} \left( \Re, \phi, \{ \br_{k=1}, \dots,\br_{k = N_p - 1} \} \right),\\
		\Delta \bT_{j \to i} &= \bm{g}_{j \to i} \left( \Re, \phi, \{ \br_{k=1}, \dots,\br_{k = N_p - 1} \} \right).
	\end{align}
\end{subequations}
\subsection{Pairwise interactions assumption}\label{sec:pairwise}
\begin{figure}[t]
	\centering
	\includegraphics[width=0.3\textwidth]{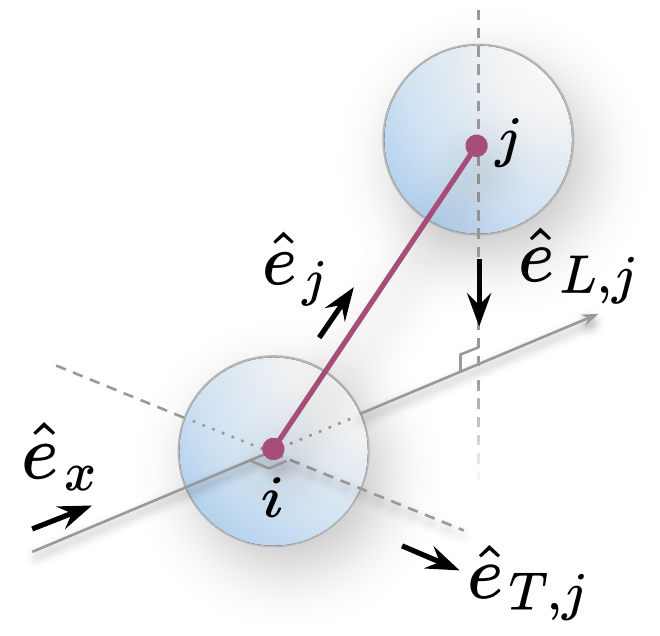}
	\caption{Visualization of the vectors along which the streamwise (i.e. drag) and lateral force (i.e. lift) and lateral torque influences due to the neighboring particle $ j $ on the reference particle $ i $ are directed under the pairwise interactions assumption.}
	\label{fig:vectors}
\end{figure}

The general relations above are overwhelmingly complex and thus formidably difficult, if not impossible, to directly determine for dense, inertial particle-laden flows. An effective approximation that has turned out to yield reasonably accurate outcomes both for the PIEP \cite{Akiki_2017,Moore2019,Balachandar2020} and MPP \cite{SeyedAhmadi2020} models is to assume pairwise hydrodynamic interactions. Having specified $ \Re $ and $ \phi $ for a certain flow regime, this approximation takes the influence of each individual neighbor $ j $ on the force or torque of particle $ i $ to be solely dependent on the location of the neighbor $ j $. With this major simplification, \cref{eq:general_form2} becomes
\begin{subequations}
	\label{eq:pairwise}
	\begin{align}
		\Delta \bF_{j \to i} &\approx \bm{f}_{j \to i} ( \br_{j} ),\\
		\Delta \bT_{j \to i} &\approx \bm{g}_{j \to i} ( \br_{j} ).
	\end{align}
\end{subequations}
The other important implication of the pairwise interactions assumption is that the flow disturbance created by a single neighbor around particle $ i $ will be symmetric about the plane that contains $ \hat{\be}_x $ and $ \hat{\be}_j $. As a result, streamwise and lateral components of each force influence $ \bm{f}_{j \to i} ( \br_{j} ) $ cannot have a component normal to that plane, and hence must lie in the same geometrical plane as the one formed by the vectors $ \hat{\be}_x $ and $ \hat{\be}_j $. The force influence $ \bm{f}_{j \to i} ( \br_{j} ) $ can therefore be decomposed as
\begin{align}
	\label{f_ij}
	\bm{f}_{j \to i} ( \br_{j} ) &= \Delta D_j(\br_j) \hat{\be}_x + \Delta L_j(\br_j) \hat{\be}_{L, j},
\end{align}
where
\begin{align}
	\label{eq:e_L}
	\hat{\be}_{L, j} = \frac{\hat{\be}_x \times (\hat{\be}_x \times \hat{\be}_{j})}{\| \hat{\be}_x \times (\hat{\be}_x \times \hat{\be}_{j}) \|}.
\end{align}
As for the torque influence, $ \Delta \bT_{j \to i} $ will neither have a streamwise nor a lateral component along $ \hat{\be}_{L, j} $ under the assumption of pairwise interactions, due to the same flow symmetry argument. Therefore, $ \Delta \bT_{j \to i} \cdot \hat{\be}_L = 0 $ and $ \Delta \bT_{j \to i} \cdot \hat{\be}_x = 0 $, and the torque influence can be written as
\begin{align}
	\label{g_ij}
	\bm{g}_{j \to i} ( \br_{j} ) &= \Delta T_j(\br_j) \hat{\be}_{T, j},
\end{align}
where the direction of the lateral component is similarly dictated by the disturbance flow symmetry and is given as
\begin{align}
	\label{eq:e_T}
	\hat{\be}_{T, j} = \frac{\hat{\be}_x \times \hat{\be}_{j}}{\| \hat{\be}_x \times \hat{\be}_{j} \|}.
\end{align}
A depiction of the relevant vectors in a pairwise configuration is given in \cref{fig:vectors}.
\subsection{Unified function representation}
In \cref{f_ij,g_ij}, the unknown functions $ \Delta D_j(\br_j) $, $ \Delta L_j(\br_j) $ and $ \Delta T_j(\br_j) $ are assumed to be distinct functions for each neighbor $ j $. However, such a dependence on the neighbor order may not be crucial or even necessary for a successful modeling effort. One reason is that since each function is dedicated to a different neighbor, the range of their inputs (i.e. $ \br_{j} $s) would be different and minimally overlapping. This suggests the possibility of using a single function that handles all $ \br_{j} $s regardless of their particular order. In the context of regression, this would be equivalent to using a global regression function for the entire range of values instead of using piecewise functions for each particular range.
The other reason originates from our observations of probability distribution maps (PDFs) pertaining to different neighbors in \cite{SeyedAhmadi2020}. We found that despite the PDFs expectedly covering different regions of space around the reference particle owing to their different distancing, the overall qualitative patterns were strikingly similar. This was why we were able to use global PDFs for all neighbors without any remarkable loss of accuracy in the model compared to using separate PDFs for each neighbor.

The unified function representation approach is similar, but not identical to the \textit{order invariance} approximation employed by Moore et al. \cite{Moore2019}. In \cite{Moore2019}, the dependency of force and torque influence functions on the neighbor number is removed, and a PDF inferred from the training data is used as a weighting function to account for the differences in the strengths of effects due to different neighbors. In the proposed model here, the ordering of the neighbors is implicitly accounted for by the approximating influence functions. The problem hence reduces to the estimation of the three unknown scalar functions $ \Delta D(\br_j) $, $ \Delta L(\br_j) $ and $ \Delta T(\br_j) $. Consequently, the final form of the total streamwise and lateral force and torque contributions to deviations from average values can be written as
\begin{subequations}
	\label{eq:F_T_final}
	\begin{align}
		\Delta \bF_i &= \sum_{j=1}^{N_p - 1} \Delta D(\br_j) \hat{\be}_x + \Delta L(\br_j) \hat{\be}_{L, j},\\
		\Delta \bT_i &= \sum_{j=1}^{N_p - 1} \Delta T(\br_j) \hat{\be}_{T, j},
	\end{align}
\end{subequations}
with $ \hat{\be}_{L, j} $ and $ \hat{\be}_{T, j} $ given in \cref{eq:e_L,eq:e_T}.
\section{Neural network model}
\begin{figure}[t]
	\centering
	\begin{subfigure}[c]{0.6\textwidth}
		\caption{}
		\centering
		\includegraphics[width=\textwidth]{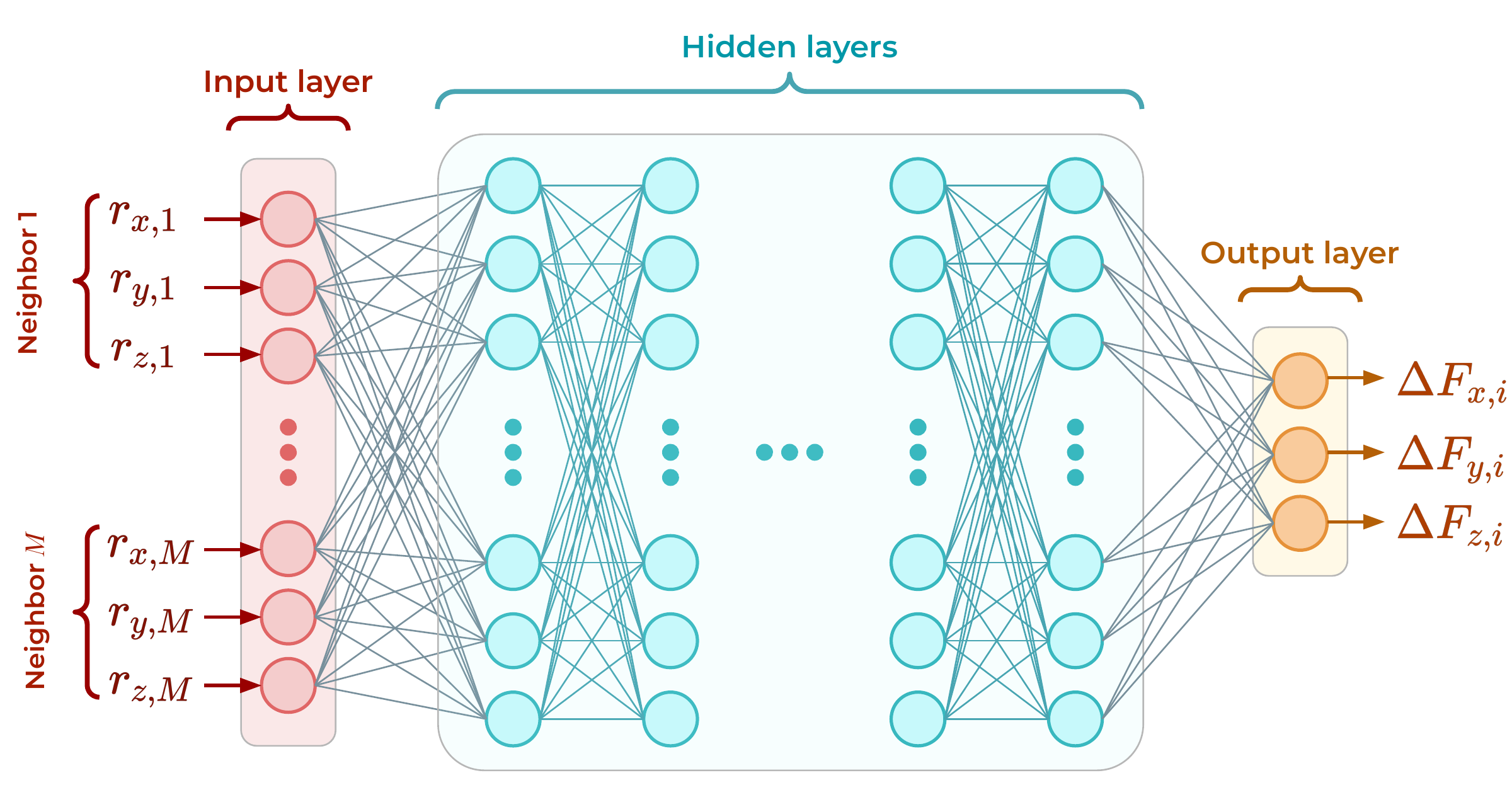}
		\label{fig:fully_connected_NN}
	\end{subfigure}
	\quad
	\begin{subfigure}[c]{0.25\textwidth}
		\caption{}
		\centering
		\includegraphics[width=\textwidth]{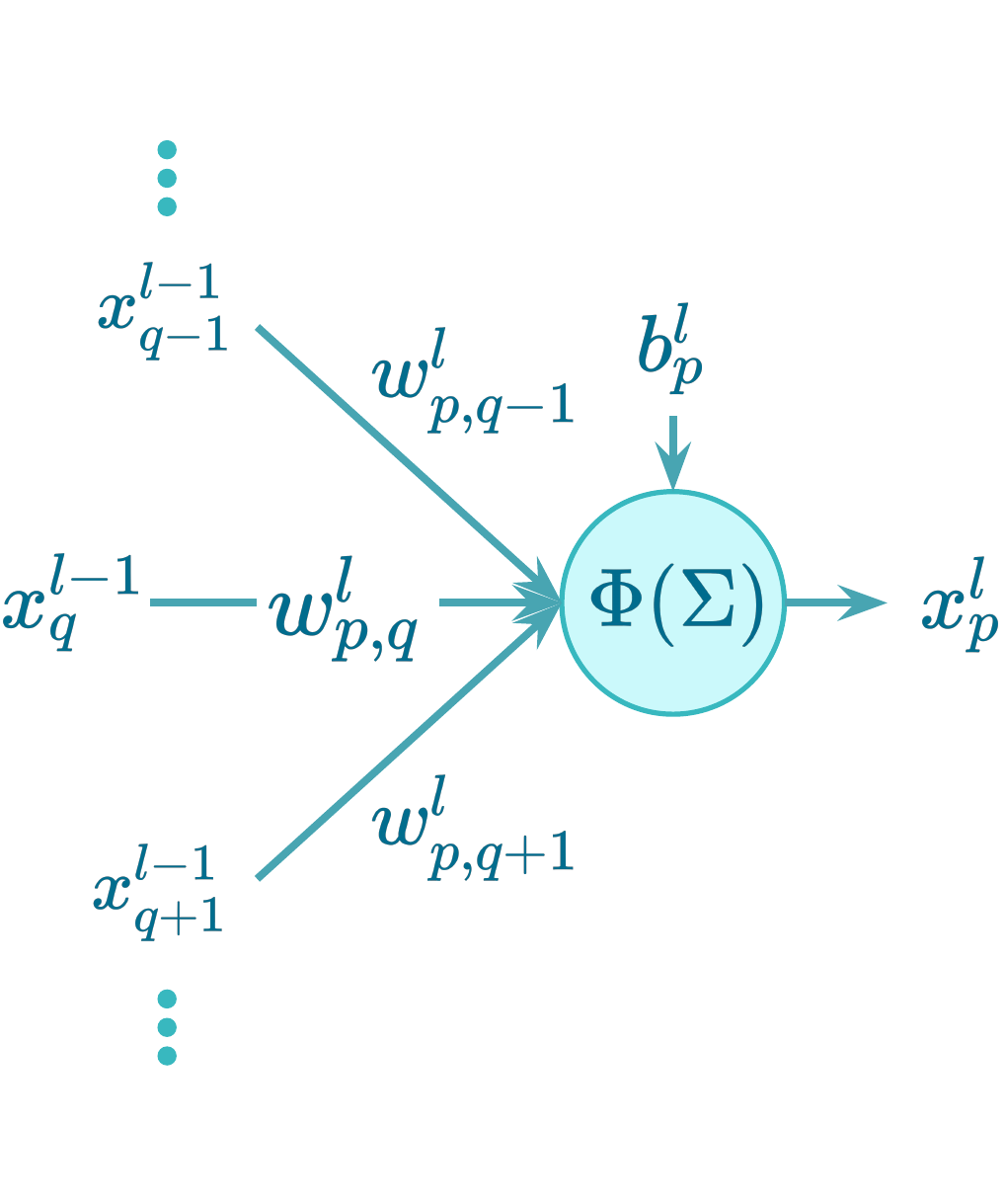}
		\label{fig:neuron}
	\end{subfigure}
	\caption{\textbf{(a)} Structure of a typical fully connected neural network model, \textbf{(b)} the schematic view of a single neuron}
	\label{fig:NN}
\end{figure}
Artificial NNs are arguably the most widely used machine learning method, known for their potential to approximate any arbitrarily complex nonlinear function \cite{Hornik1989}. A schematic view of a typical NN structure is shown in \cref{fig:NN}. The main component of NNs are nodal processing units called ``neurons'', which are conceptually inspired by their biological counterparts, but with no direct correspondence of any kind. The neurons are usually stacked up in the form of sequential layers as can be seen in \cref{fig:fully_connected_NN}. The lines between pairs of neurons represent connections with unique weighting parameters associated with them, shown as $ w^l_{p,q} $ in \cref{fig:neuron}.
Each neuron in layer $ l $ receives inputs $ x^{l-1}_q $ from neurons in the preceding layer $ l-1 $, and computes a weighted sum of the incoming data. A bias term $ b^l_p $ is then added to the sum, and the resulting value is passed through an activation function $ \Phi $, the nonlinearity of which enables the NN to approximate nonlinear mappings. The outcome of operations done by neuron $ p $ in layer $ l $, i.e. $ x^l_p $ is given as
\begin{align}
	\label{eq:x_l_p}
	x^l_p &= \Phi \left( \sum_q w^l_{p,q} x^{l-1}_q + b^l_p \right).
\end{align}
The training phase of a NN is the process through which optimal values for the weights and biases are found. The process begins with random (or in some cases prescribed) initialization of weights and biases in the first iteration. Raw data is then taken via the input layer, where the information merely gets relayed to the next layer without any operations being performed. The data undergoes successive transformations in the intermediate layers---also known as hidden layers---until it reaches the output layer, thus completing a single feed-forward step. The predicted values generated by the NN are then compared with the expected target values in the dataset by a so-called loss or cost function. The actual training of the network takes place in the back-propagation step \cite{Rumelhart1986}, where the gradient of the loss function with respect to the weights and biases of the network are efficiently computed using the chain rule. These gradients are utilized by an optimization algorithm, e.g. gradient descent or its variants, to adjust the weights and biases such that the loss function is minimized.
Owing to their highly flexible nature, NNs are prone to overfitting and lack of generalizability. A model that suffers from overfitting exhibits high prediction power on the training dataset, but performs poorly when presented with unseen data. In such a case, the model is complex enough that it merely memorizes the training dataset without identifying the true underlying patterns. It is common practice in training NNs to split the dataset into training and validation sets in order to be able to detect overfitting. The model is trained using the former, while tested on the latter to ensure similar performance on both sets. Various regularization strategies that control model complexity exist to prevent overfitting, such as parameter norm penalties and the dropout technique \cite{Goodfellow2016}. A NN model may also be regularized by devising particular architectures correspondent to the target problem to be solved, or by imposing problem-specific cost functions that ensure the satisfaction of governing laws and constraints. This matter will be discussed further in the subsequent sections.

For the problem of interest in this work, the input data consists of the components of position vectors pertaining to $ M $ influential neighbors that surround a test particle (shown as $ \{ r_{x,j}, r_{y,j}, r_{z,j}\} $ for each neighbor $ j $ in \cref{fig:fully_connected_NN}), whereas the outputs are either hydrodynamic force or torque components (shown in \cref{fig:fully_connected_NN} only as $ \{ \Delta F_{x,i}, \Delta F_{y,i}, \Delta F_{z,i}\} $ for brevity). For a NN that is to supposed to perform regression, a standard choice for the loss function is the mean squared error (MSE) of the predicted output with respect to the expected value. For a given dataset and network configuration, the loss function based on the MSE is defined as
\begin{align}
	\label{eq:loss}
	\bm{L}(\bm{w}, \bm{b}) &= \frac{1}{N_p} \sum_i \left( \Delta \bF_{i,\text{NN}} - \Delta \bF_{i,\text{DNS}} \right)^2,
\end{align}
where $ \bm{w} $ and $ \bm{b} $ denote the weights and biases matrices, respectively. Note that for a model with multiple outputs, the final loss value may be obtained by summing individual losses for each component in $ \bm{L} $.
\subsection{Fully connected NN architecture}\label{sec:FCNN}
The most basic type of NNs is one that has multiple layers of neurons stacked sequentially between the input and output layers. The characteristic feature of such NNs is that each node within the network is connected to each and every other adjacent node as shown in \cref{fig:fully_connected_NN}, hence explaining the origin of the term \textit{fully-} or \textit{densely-connected} used to describe these NN structures.
From the modeling perspective, the problem of force and torque prediction based on local neighborhood is a fairly complex one: the underlying mapping function is expected to be nonlinear and it typically requires the positions of $ \approx 10-20 $ neighbors as input for accurate approximations \cite{Moore2019,SeyedAhmadi2020}. For such a high-dimensional problem governed by potentially complex nonlinear functions, a FCNN needs to be sufficiently flexible to be able to model the intricacies of the target function.
Even though the desired flexibility can be readily achieved by deepening and widening the network (i.e. increasing the number and the width of the hidden layers), successful training becomes seriously impeded in this scenario. Balachandar et al. \cite{Balachandar2020} have recently shown that a FCNN cannot be directly trained on limited PR-DNS data for accurate prediction of hydrodynamic forces. This is consistent with and confirmed by our experiments with FCNNs, as will be later discussed in \cref{sec:results}.
With the growing number of training parameters (i.e. weights and biases), a limited dataset generated from expensive PR-DNS becomes insufficient for effective learning of generalizable patterns in the data, often resulting in either inability to minimize the loss function altogether or severe overfitting. Moreover, as the input space dimensions increase, exponentially more data is required to maintain a constant data density due to the \textit{curse of dimensionality} \cite{Hastie2009}.

Overcoming the foregoing challenges requires special designing of the NN structure and tailoring it to the particular problem at hand. In fact, a great part of the success of NNs in general domains is owed to the specialized architectures of these models for particular tasks. Remarkable examples are convolutional NNs which are capable of handling high-dimensional data for image recognition \cite{NIPS2012_4824}, whereas long short-term memory NNs are designed for modeling sequential data such as time series \cite{Hochreiter1997}.
When physical systems are concerned, domain knowledge needs to be incorporated in the way an ML model is constructed and trained to ensure that physical constraints and governing laws are properly enforced. Such prior knowledge can be embedded in ML models in the form of observational, structural and learning biases \cite{Karniadakis2021}. Promising examples in fluid mechanics are implicit enforcement of invariances and symmetry \cite{Ling2016a} especially in turbulence modeling \cite{Wu2018,Ling2016}, or direct embedding of governing equations as regularization mechanisms for penalizing the loss function \cite{Raissi2019,Raissi2018}.

\subsection{Physics-inspired NN architecture}\label{sec:PINN}
\begin{figure}[t]
	\centering
	\includegraphics[width=1\textwidth]{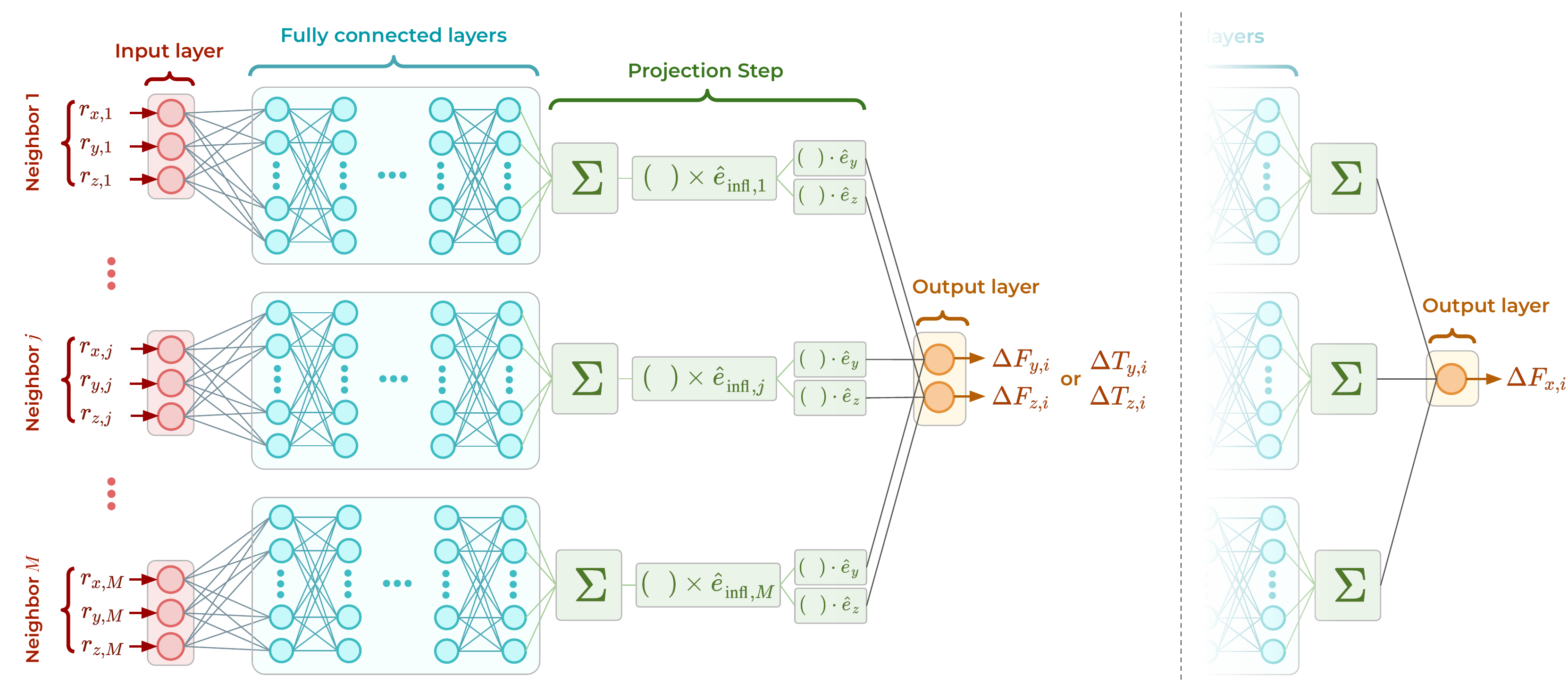}
	\caption{A schematic view of the architecture of the proposed physics-inspired neural network model. The diagram to the left of the dashed vertical line shows the model design when lift or torque is predicted. When prediction of drag is desired, the projection step is trivial and hence unnecessary, leading to a simplified structure shown to the right of the vertical dashed line.}
	\label{fig:PINN}
\end{figure}
In the present work, we propose a specialized NN architecture that incorporates the simplifying approximations presented in \cref{sec:model_formulation}; namely, the pairwise interactions and unified function representation assumptions. Strictly speaking, the physics-inspired neural network (PINN) specifically represents \cref{eq:F_T_final}, the structure of which is depicted in \cref{fig:PINN}. The PINN implements the two model approximations as follows:
1) With this architecture, instead of simultaneously feeding the relative positions of all influential neighbors to a single FCNN, separate NN blocks are responsible for handling the influences of individual neighbors. These influences are then summed up to give the total effect of the neighborhood of a particle.
2) The training parameters (weights and biases) are shared between the NN blocks (shown in light blue in \cref{fig:PINN}), constraining them to approximate a single unified function for all neighbors.
Consequently, the flow of data in the PINN is as follows:
\begin{enumerate}[label=\roman*)]
	\item The components of the position vector of each neighbor $ j $, i.e. $ r_{x,j} $, $ r_{y,j} $ and $ r_{z,j} $ are fed to its corresponding NN block.

	\item The information is processed in the NN blocks through multiple layers of densely connected neurons, the parameters of which are shared among them. A single scalar output is generated by summing the outputs of the last layer to represent each of the three functions $ \Delta D(\br_j) $, $ \Delta L(\br_j) $ and $ \Delta T(\br_j) $ in \cref{eq:F_T_final}.

	\item The resulting scalar values generated in the preceding step are then multiplied by appropriate direction vectors; namely, $ \hat{\be}_x $, $ \hat{\be}_{L,j} $ and $ \hat{\be}_{T,j} $ for the streamwise and lateral forces (i.e. drag and lift, respectively) as well as the lateral torque. This is done to generate the force and torque influence vectors of \cref{eq:F_T_final}, each of which is then projected onto the $ x $, $ y $ and $ z $ directions.

	\item The output layer is set to have a linear activation function and no bias parameter. This means that this layer is merely responsible for computing a weighted sum of the force/torque influences that it receives from the network pathway of each neighbor, as shown in the diagram in \cref{fig:PINN}. This completes one feed-forward pass of the PINN.
\end{enumerate}
In the training phase, the outputs generated in the last step above are compared with the PR-DNS values, and back-propagation scheme adjusts the parameters of the PINN using the gradients of the loss function. The feed-forward and back-propagation steps are iteratively repeated until stopping conditions (e.g. maximum number of iterations) are met.

The PINN structure described above is deliberately designed to introduce significant structural bias into the model, corresponding directly to the physical formulation of the problem. Specifically, the proposed architecture reduces the potential NN modeling complexity in two distinct ways.
First, each NN block always handles only $ 3 $ input variables pertaining to a single neighboring particle, regardless of the number of included neighbors in the model. Since the pairwise interactions assumption ignores second- and higher-order interactions, the PINN mapping function that represents such a simplification becomes substantially less complex, and thus more amenable to modeling.
Second, all NN blocks share the same set of parameters, meaning that a single unified influence function is approximated for all neighbors regardless of their distance. This feature is reminiscent of parameter sharing of convolving filters in CNNs, which is a crucial technique for complexity reduction.
Consequently, while the former strategy simplifies the nature of the mapping function itself, the latter reduces the number of such mappings to unity on justified grounds. As we will see in the next section, the proposed PINN demonstrates strong performance for all considered cases. This will be contrasted with the FCNN model that fails to provide accurate predictions of forces and torques in most scenarios and ultimately suffers from overfitting.

\subsection{Influence of local average velocity}
We pointed out in \cite{SeyedAhmadi2020} that even though the immediate neighborhood of a particle determines the major part of the force and torque variance of individual particles, not every group of particles is exposed to the same mesoscale flow.
Consider a particular arrangement of particles that in one instance is fully exposed to a channel of flow, while another instance of the same arrangement is partly shielded from the flow by an upstream group of particles. In this scenario, the undisturbed flow seen by a particle in the former group would be different than that seen by a particle in the latter group, despite the arrangement being identical.
Obviously, the average macroscale flow is the same for all particles only on the scale of the entire bed, which means that capturing this effect requires the inclusion of the influence of all particles in the bed for predicting force and torque of each particle. Our present model cannot capture this effect directly even if all $ N_p - 1 $ neighbors of each particle were included in the training process. Modeling such a highly nonlinear, high dimensional functional dependence goes beyond the underlying assumptions of the proposed PINN model.
Consequently, we adopt the same strategy as in \cite{SeyedAhmadi2020} so as to partly capture the foregoing effects. To this end, we compute the local volume-averaged velocity around each particle within the bed, and add that as an additional predictive input feature to the NN models (see \cite{SeyedAhmadi2020} for details of the averaging process). In terms of the NN structure, this is achieved by supplying the average velocity components as scalar inputs to the output layer of the NN models in \cref{fig:NN,fig:PINN} with linear activation functions. Similar to the findings of \cite{SeyedAhmadi2020}, improvement in performance appears to be most pronounced in case of drag prediction, with the increasing of $ \Rsq $ ranging from $ \Delta \Rsq \approx 0.05 $ to $ 0.15 $ for $ \phi = 0.1 $ and $ \phi = 0.4 $, respectively.

\subsection{Hyper-parameters and implementation}
So far, the overall structures of both types of NN models are described in \cref{sec:FCNN,sec:PINN}. There are a number of choices that need to be made with regards to the specific configuration of each model by deciding the values of model hyper-parameters.
We have found in our experiments with the PINN model that best results are achieved with two hidden layers for the NN blocks, each consisting of 10-15 neurons. In case of the FCNN model, no significant improvement was observed between having one or more hidden layer. The addition of more layers in the case of FCNN leads to the faster deterioration of the FCNN model's predictions. Moreover, the computationally efficient Adam optimizer \cite{kingma2017adam} with a learning rate between $ 0.001 $ and $ 0.01 $ is used for the minimization process. The maximum number of allowable epochs (i.e. one complete pass through the entire dataset) is set to $ 2000 $, with a batch size of $ 100 $.
Furthermore, among various existing choices for the activation function, the hyperbolic tangent seems to be preferable for most regression models and is therefore employed for the present work. Finally, we use the $ k $-fold cross-validation technique \cite{Hastie2009} to monitor the model for possible overfitting. The $ k $-fold method works by first partitioning the dataset into $ k $ segments or folds. The model is then iteratively validated with each fold while being trained on the remaining $ k-1 $ folds. In this manner, the entire dataset is both trained on and validated against, and the model becomes robust to particular choices of the training and validation sets. As for the results, we will report model performance measures that are averaged over all rounds of the cross-validation.

The NN models presented in this work are implemented using TensorFlow \cite{tensorflow2015-whitepaper}, which is Google's free and open-source library for training deep NNs. In doing so, we used the Keras \cite{chollet2015keras} functional API in Python which enabled us to employ a customized architecture for the PINN model. The codes and datasets of this work are publicly available and can found through the following link: \url{https://github.com/armanawn/PINN-force-torque-model}.
\section{Results and discussion}\label{sec:results}
In this section, we first present a performance comparison between the FCNN and PINN architectures for a few representative cases, and then report the full performance results and analysis for the developed PINN model. The performance is quantified in terms of the coefficient of determination defined as
\begin{align}\label{eq:R^2}
	\Rsq = 1 - \frac{\displaystyle\sum_{i=1}^{N} \left( \Delta F_{\text{Model}, i} - \Delta F_{\text{DNS}, i} \right)^{2}}{\displaystyle\sum_{i=1}^{N} \left( \Delta F_{\text{DNS}, i} - \langle \Delta F_{\text{DNS}, i} \rangle \right)^{2}},
\end{align}
which gives the ratio of the variance explained by the model to the total variance that exist in the data. Also, $ N $ denotes the number of samples used to obtain the value of $ \Rsq $. According to \cref{eq:R^2}, $ \Rsq = 0 $ means that the model is entirely incapable of explaining variations from the mean, which corresponds to a model that merely predicts the average values of the target variable (e.g. classical drag correlations). On the other hand, a model that is able to perfectly predict the variations of the target variable would yield $ \Rsq = 1 $.
\subsection{Performance comparison of FCNN and PINN models}
\begin{figure}[t]
	\centering
	\begin{subfigure}[t]{0.3\textwidth}
		\caption{}
		\vspace{2pt}
		\includegraphics[width=0.2\textheight]{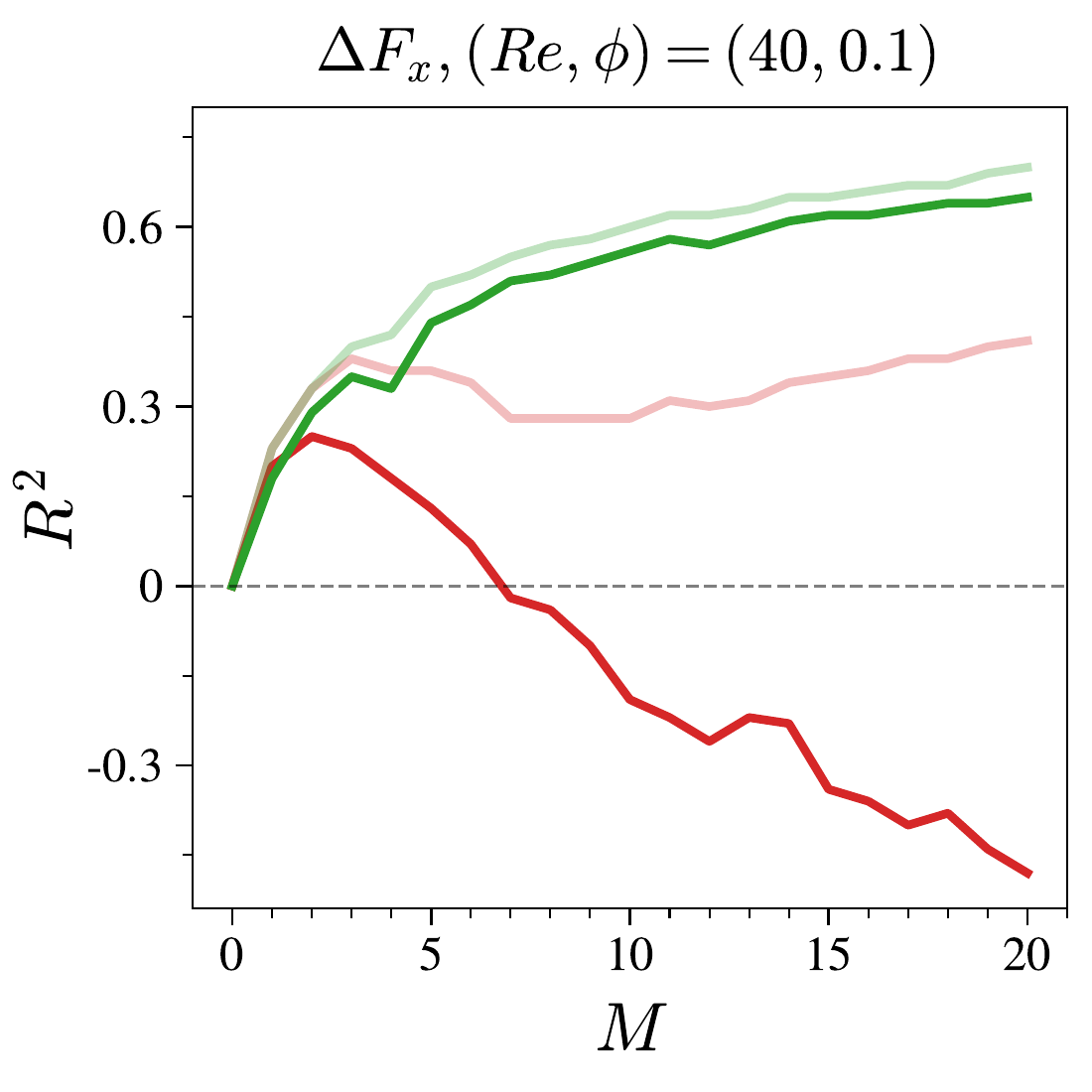}
		\label{fig:Rsq_dF_x}
	\end{subfigure}
	\quad\quad
	\begin{subfigure}[t]{0.3\textwidth}
		\caption{}
		\vspace{2pt}
		\includegraphics[width=0.2\textheight]{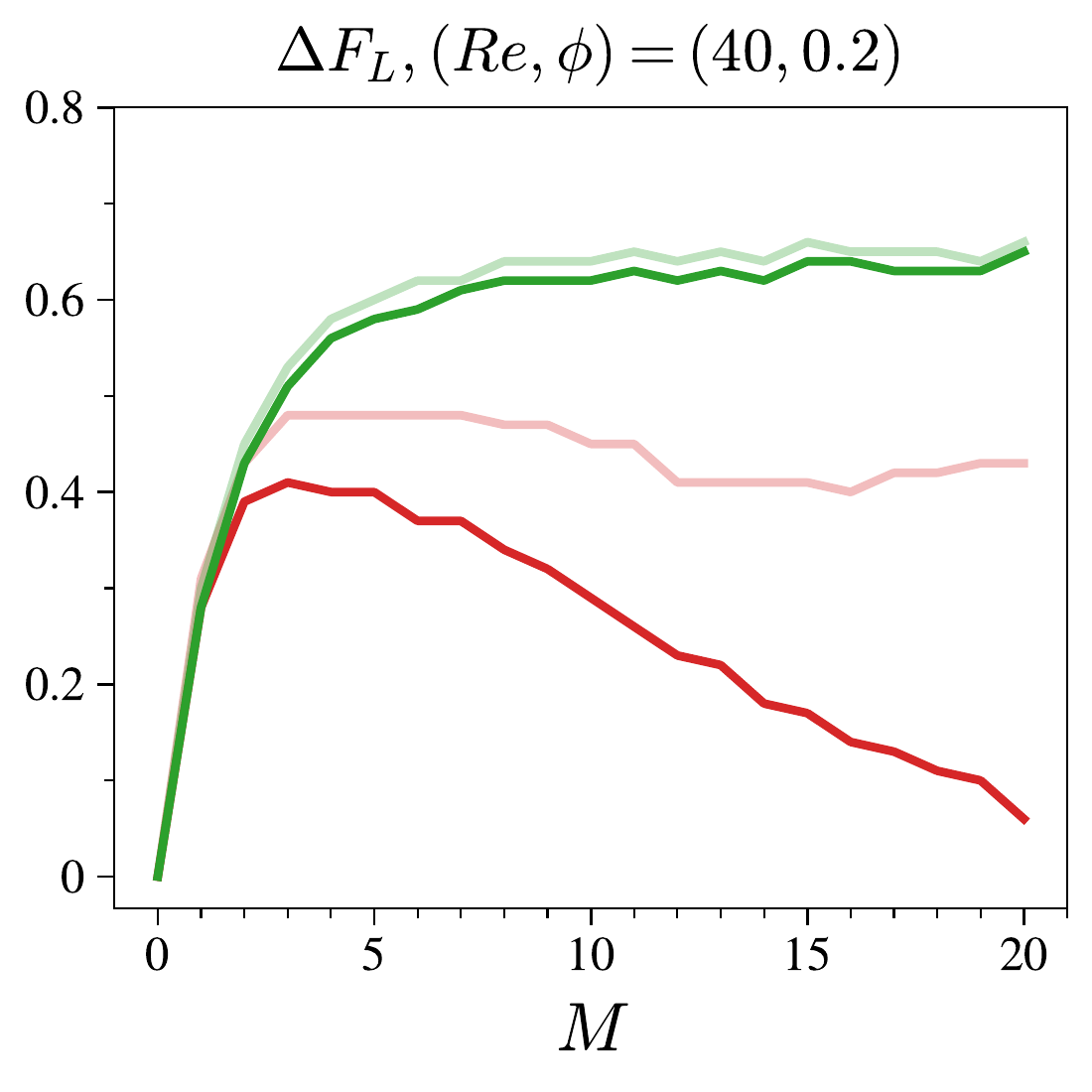}
		\label{fig:Rsq_dF_L}
	\end{subfigure}
	\quad
	\begin{subfigure}[t]{0.3\textwidth}
		\caption{}
		\vspace{2pt}
		\includegraphics[width=0.2\textheight]{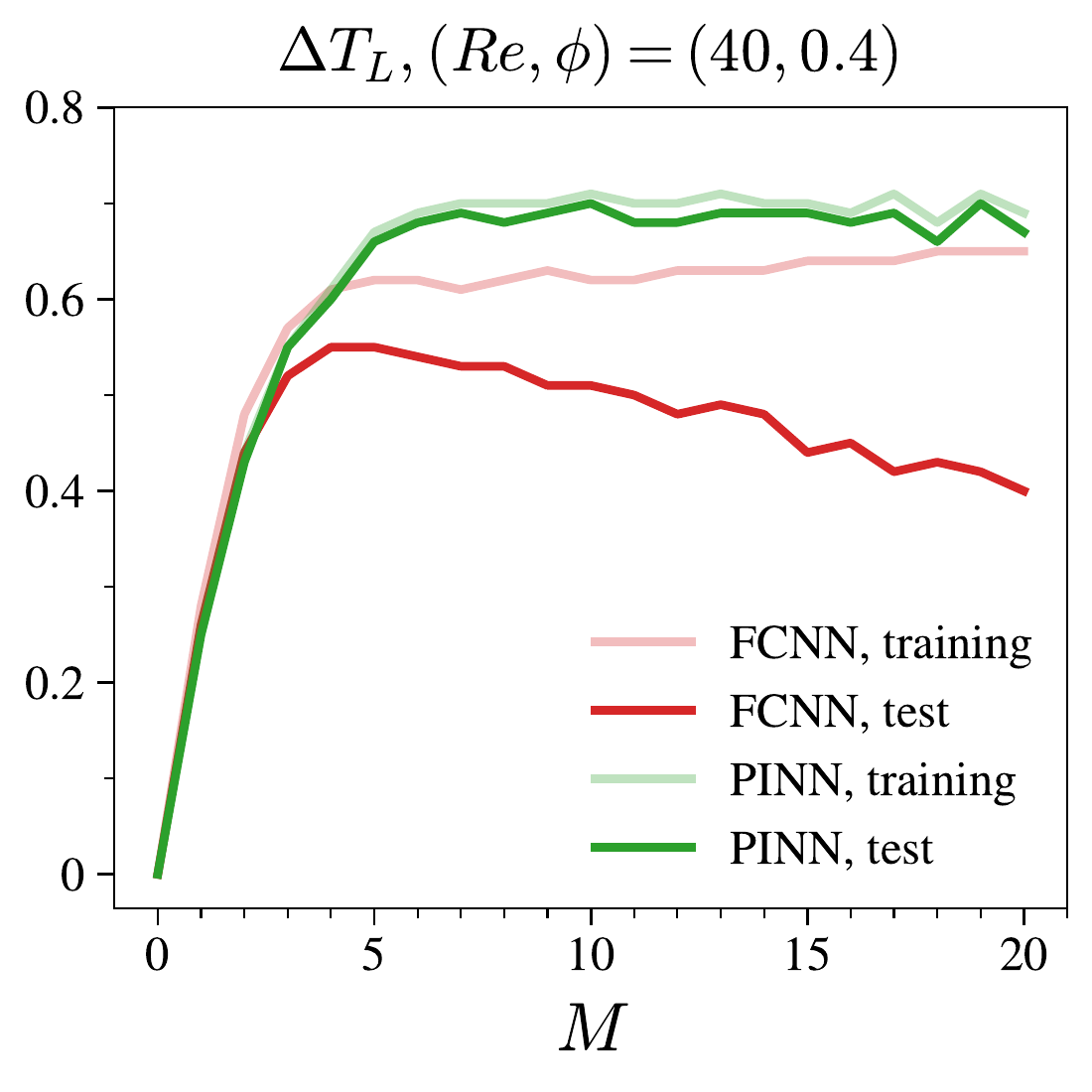}
		\label{fig:Rsq_dT_L}
	\end{subfigure}
	\caption{The performance of the PINN and FCNN models in predicting drag, lift and torque in terms of the coefficient of determination, $ \Rsq $ for different values of $ M $, i.e. the number of included neighbors in the training process. The Reynolds number in all cases is fixed at $ \Re = 40 $, whereas the solid volume fraction takes on the values of $ \phi = 0.1 $, $ 0.2 $ and $ 0.4 $.}
	\label{fig:Rsq_M}
\end{figure}
We have previously mentioned that FCNNs cannot be trained on small datasets in a direct manner to accurately predict hydrodynamic forces and torques, due to the high dimensionality and the strong nonlinearities of the problem. In \cref{fig:Rsq_M}, the coefficient of determination is shown as a function of the number of included neighbors. For the sake of conciseness, the modeling accuracy for drag, lift and torque is demonstrated only for $ \Re = 40 $ and various solid volume fractions, but our observations and conclusions also apply to $ \Re = 2 $,  $ 10 $ and $ 150 $.
We can see throughout all graphs how FCNNs fail to provide generalizable predictions as the number of included neighbors increases, whereas the accuracy of the PINN predictions improves with the addition of more neighbors to the modeling process. Notably, inclusion of more neighbor influences never leads to overfitting, but rather to an ultimate saturation of accuracy. This is the advantageous consequence of parameters sharing between NN blocks that decouples the number of trainable parameters in the model from the number of included neighbors. \Cref{fig:Rsq_dF_x} shows that while the coefficient of determination for the test set steadily increases and reaches $ \Rsq \approx 0.71 $ for drag in case of the PINN model, the FCNN model peaks at $ \Rsq \approx 0.28 $ with the inclusion of $ M = 2 $ neighbors. The lines shown in red that indicate the training and test set performance of the FCNN model begin to diverge at this point. The $ \Rsq $ associated with the test set experiences an almost monotonic drop beyond this point, showing a characteristic manifestation of overfitting. In the case of drag, the $ \Rsq $ of the FCNN drops below zero for $ M \ge 7 $, meaning that the mean drag value is a better prediction of the drag compared to the output of the FCNN model. Balachandar et al \cite{Balachandar2020} also found negative values of $ \Rsq $ for the prediction of drag with $ M = 25 $; however, they do not report $ \Rsq $ for smaller $ M $.
The scenario is qualitatively the same for the prediction of the lateral force and torque, as can be seen in \cref{fig:Rsq_dF_L,fig:Rsq_dT_L}, respectively. Quantitatively, we find that the FCNN model performs better in predicting the lateral loads compared to the drag, attaining a maximum $ \Rsq $ of $ \approx 0.4 $ and $ \approx 0.5 $ for lift and torque with $ M = 3 $ and $ 4 $, respectively. Nevertheless, the PINN model consistently achieves $ 50\% $ to $ 70\% $ higher $ \Rsq $ in comparison with the FCNN model, with no sign of overfitting.

The behavior of the FCNN and PINN models in terms of their performance versus the number of included neighbors for drag, lift and torque merits further elaboration. Regardless of the case, we have seen that the FCNN model always suffers from overfitting when $ M $ is increased beyond a certain number. Unlike the PINN, the FCNN model does not assume any particular structure for the functional dependence of the hydrodynamic forces and torques on the local neighborhood. As more neighboring particles are included in the model, the FCNN needs to become increasingly more intricate so as to be capable of handling complex interactions between the input variables. This can be achieved by employing more hidden layers, and more neurons in each layer. Despite being theoretically flexible enough to approximate any arbitrarily complex functional dependence, training a FCNN of such complexity on a small dataset remains unfeasible in practice. Consequently, the added complexity benefits an elaborate FCNN only to memorize the given dataset instead of learning generalizable patterns. The overfitting behavior observed in \cref{fig:Rsq_M} is a manifestation of the foregoing discussion.
The PINN model, on the other hand, seems to be immune to overfitting thanks to a successful regularization strategy for controlling model complexity; namely, its structural bias. In this case, the physical formulation of the problem (see \cref{sec:model_formulation}) incorporated in the architecture of the PINN effectively guides the training procedure. Nevertheless, a clear upper bound for the performance of the PINN model is evident in \cref{fig:Rsq_M}. In contrast to the case of a FCNN, this seeming saturation of performance of the PINN is a theoretical limit rather than a practical one, as for the FCNN. The proposed structure of the PINN follows a physical formulation which assumes binary hydrodynamic interactions between the particles. This means that the particular formulation presented in \cref{sec:model_formulation} only considers direct, first-order influences of the neighboring particles on the force or torque of a reference particle, and neglects any indirect, higher-order interactions. In this sense, the performance limit that we observe is not surprising, but expected indeed.
\subsection{Overall performance of the PINN model}
\begin{figure}[t]
	\centering
	\begin{subfigure}[t]{0.3\textwidth}
		\caption{}
		\vspace{2pt}
		\includegraphics[width=0.95\textwidth]{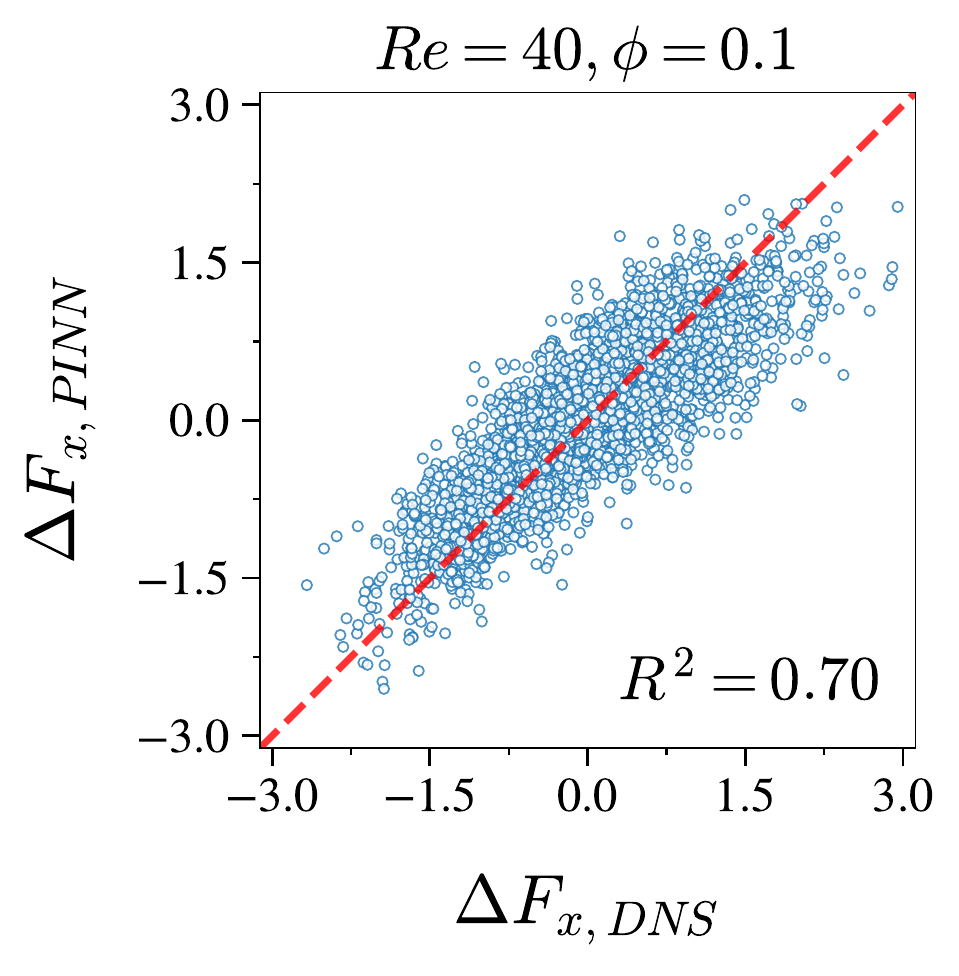}
		\label{fig:corr_dF_x}
	\end{subfigure}
	\begin{subfigure}[t]{0.3\textwidth}
		\caption{}
		\vspace{2pt}
		\includegraphics[width=0.95\textwidth]{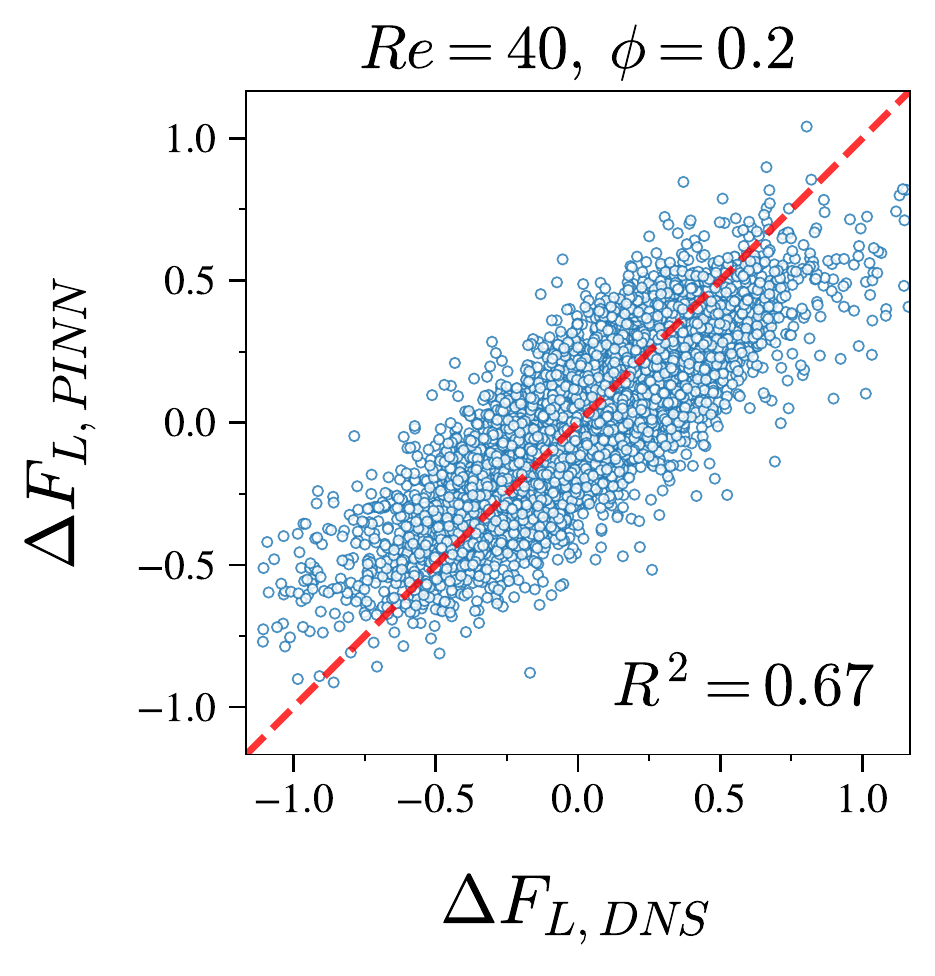}
		\label{fig:corr_dF_L}
	\end{subfigure}
	\begin{subfigure}[t]{0.3\textwidth}
		\caption{}
		\vspace{2pt}
		\includegraphics[width=0.95\textwidth]{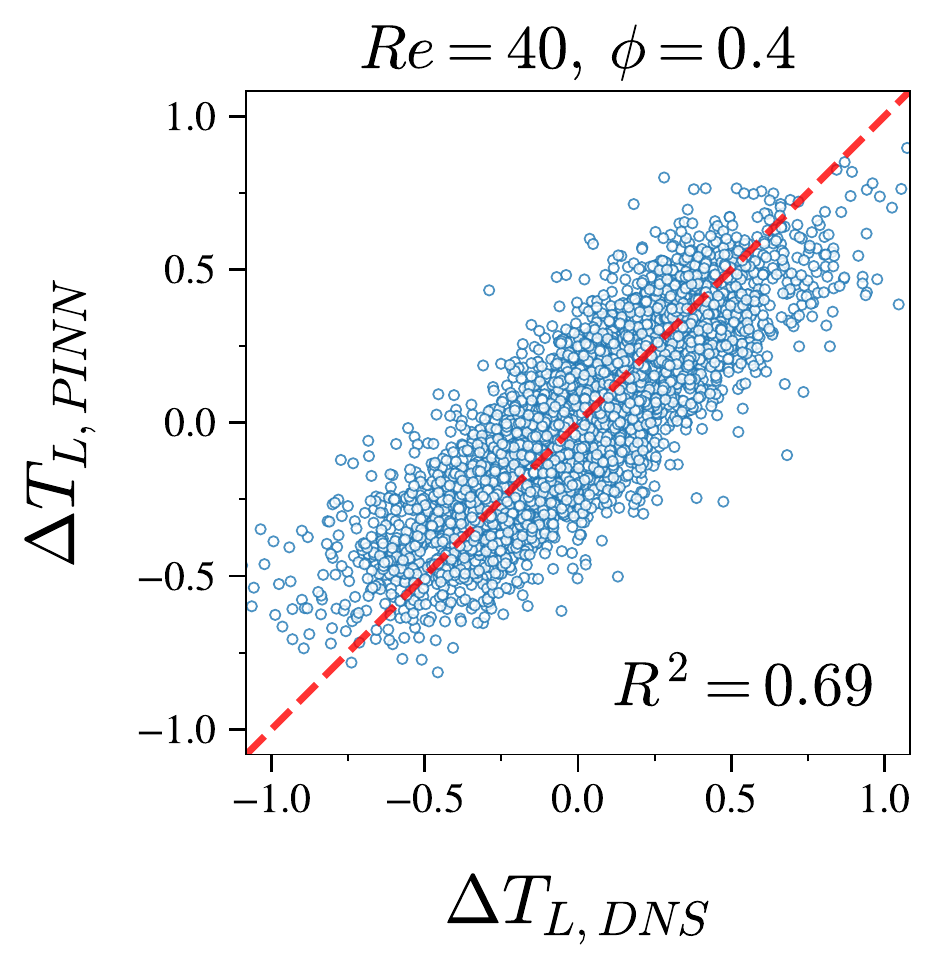}
		\label{fig:corr_dT_L}
	\end{subfigure}
	\caption{Correlation plots generated after training the PINN model to predict drag, lift and torque at $ \Re = 40 $ and various solid volume fractions. The coefficient of determination, $ \Rsq $ for each case is also given. The red dashed line represents an ideal fit.}
	\label{fig:corr_plots}
\end{figure}

\newcommand{\grrr}{\cellcolor{green!25}}
\newcommand{\grr}{\cellcolor{green!15}}
\newcommand{\gr}{\cellcolor{green!5}}
\newcommand{\rr}{\cellcolor{red!5}}
\begin{table}
	\centering
	\setlength{\tabcolsep}{20pt}
	\begin{tabular}[c]{C C C C C}
		\toprule
		\phi	&	\Re	&	\Delta F_x	&	\Delta F_L  & \Delta T_L\\
		\midrule
		0.1 &	2	&	\grrr{0.81}	& \grrr{0.77}	&	\grrr{0.89}\\
		0.1 &	10	&	\grrr{0.82}	& \grrr{0.78}	&	\grrr{0.86}\\
		0.1 &	40	&	\grrr{0.70}	& \grrr{0.73}	&	\grrr{0.74}\\
		0.1 &	150	&	\grr{0.61}	& \grr{0.66}	&	\grr{0.65}\\
		\\
		0.2 &	0.2	&	\grr{0.68}	& \gr{0.57}		&	\grrr{0.74}\\
		0.2 &	2	&	\grrr{0.73}	& \grr{0.68} 	&	\grrr{0.88}\\
		0.2 &	40	&	\grrr{0.71}	& \grr{0.67}	&	\grrr{0.76}\\
		0.2 &	150	&	\grr{0.61}	& \gr{0.56} 	&	\gr{0.56}\\
		\\
		0.4 &	2	&	\grr{0.61}	& \gr{0.51}		&	\grrr{0.72}\\
		0.4 &	40	&	\grr{0.66}	& \gr{0.58}		&	\grr{0.69}\\
		0.4 &	150	&	\gr{0.52}	& \gr{0.53}		&	\gr{0.59}\\
		\bottomrule
	\end{tabular}
	\caption{Performance of the PINN model in terms of the coefficient of determination $ \Rsq $}
	\label{table:R_sq}
\end{table}
The performance of the PINN model is evaluated for all cases in \cref{table:cases} in terms of the coefficient of determination, and the results are presented in \cref{table:R_sq}. Moreover, correlation plots are also generated for three cases in \cref{fig:corr_plots}, representing modeling results for drag (\cref{fig:corr_dF_x}), lift (\cref{fig:corr_dF_L}) and torque (\cref{fig:corr_dT_L}) at $ \Re = 40 $ and various $ \phi $. The horizontal and vertical axes of each plot show the values obtained from PR-DNS (i.e. exact values) and PINN model, respectively. Each data point on the plots represents the prediction result for a single sample in the dataset. The red dashed line indicates an ideal, perfect model where the predictions of the PINN model are equal to those given by PR-DNS. Therefore, the closer the data points are to the red dashed line, the better the performance of the model.
The PINN model is evidently capable of explaining about two thirds of the total particle-to-particle variation in the drag, lift and torque values. In general, predictions are of higher accuracy for the torque, drag and lift in descending order. Moreover, higher $ \Rsq $ values are typically associated with regimes with smaller Reynolds numbers and lower solid volume fractions. It is remarkable that the range of $ \Rsq $ values and their variance between cases show striking resemblance to those obtained with the MPP model \cite{SeyedAhmadi2020}. Furthermore, the peak values of $ \Rsq $ that are obtained here and those reported for the PIEP model \cite{Akiki_2017,Moore2019} are also very similar. We can explain this by pointing to the fact that despite being derived in fundamentally different ways, all three models (i.e. PIEP, MPP and PINN) incorporate the same central assumption of pairwise interactions. Obviously, the extent to which this assumption compromises the model accuracy depends on the regime and the force or torque component to be modeled.

A number of interesting conclusions may be drawn from the presented results:
\begin{itemize}
	\item First-order effects captured by the pairwise interactions approximation explain the greater part of the total particle-to-particle variation of forces and torques, even in the most convective and densest flow regimes. Even though this assumption is strictly valid for $ \Re = 0 $, it still retains its predictive value albeit to a limited extent.

	\item Prediction of the hydrodynamic lift and torque requires the inclusion of far fewer neighbors compared to the case of drag (e.g. $ M = 4 $--$ 7 $), rendering the influences much more local. A similar observation was also made with the MPP model \cite{SeyedAhmadi2020}. This is linked to the fact that the flow is highly convective in the main flow direction. The induced wakes and the streamwise velocity deficit extend several diameters away from each particle, in contrast to the variations of the transverse velocity components as shown in \cite{Akiki_2017}. Similarly, the torque variation depends on the perturbation vorticity which also has a limited zone of influence. As a result, a high-accuracy prediction of the drag always requires at least $ 25 $ neighbor locations to be included in the modeling. Expectedly, we have found that this effect is more prominent for higher $ \Re $ and smaller $ \phi $, owing to the stronger convective effects and less frequent wake disruption by neighboring particles, respectively.

	\item The preceding point also shows more clearly why the FCNN model performs much worse in predicting drag compared to predicting either lift or torque. That is, the lateral force and torque of a particle are inherently influenced by fewer neighbors due to the dominance of the streamwise flow. Since the peak performance is attainable by fewer neighbors, the FCNN would be able to provide relatively better predictions before becoming excessively complex with the inclusion of more neighbors.
\end{itemize}
\subsection{Interpretability of the PINN model}
In addition to being generalizable, it is crucial for ML models of physical systems to be interpretable \cite{Brunton2020}. The PINN model developed in this paper achieves full interpretability, as its NN architecture represents the physical formulation of the problem (see \cref{sec:model_formulation}). In fact, the output of each NN block in \cref{fig:PINN} shows the scalar contribution of a neighboring particle to the deviation of the force or torque value of a test particle from the average, as a function of the neighbor location. In other words, the NN blocks approximate the functional form of the scalar functions $ \Delta D(\br_j) $, $ \Delta L(\br_j) $ and $ \Delta T(\br_j) $ in \cref{eq:F_T_final}. This is opposed to the case of the FCNN model, where it is not known on any level how an output is generated, or what the components of the NN model correspond to.
\begin{figure}[t]
	\centering
	\begin{subfigure}[t]{0.3\textwidth}
		\caption{}
		\vspace{2pt}
		\includegraphics[width=0.97\textwidth]{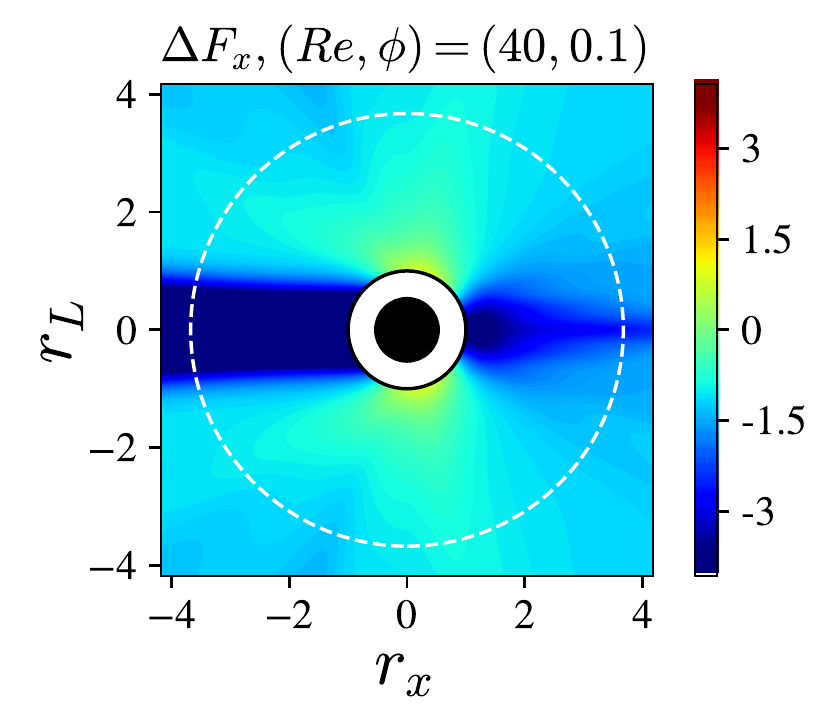}
		\label{fig:NN_out_dF_x}
	\end{subfigure}
	\begin{subfigure}[t]{0.3\textwidth}
		\caption{}
		\vspace{2pt}
		\includegraphics[width=0.95\textwidth]{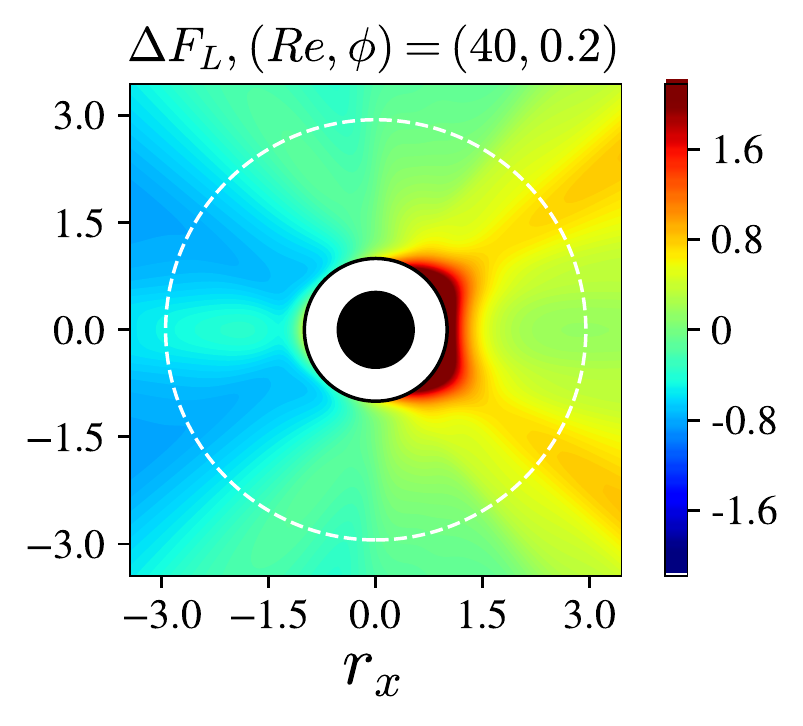}
		\label{fig:NN_out_dF_L}
	\end{subfigure}
	\begin{subfigure}[t]{0.3\textwidth}
		\caption{}
		\vspace{2pt}
		\includegraphics[width=0.95\textwidth]{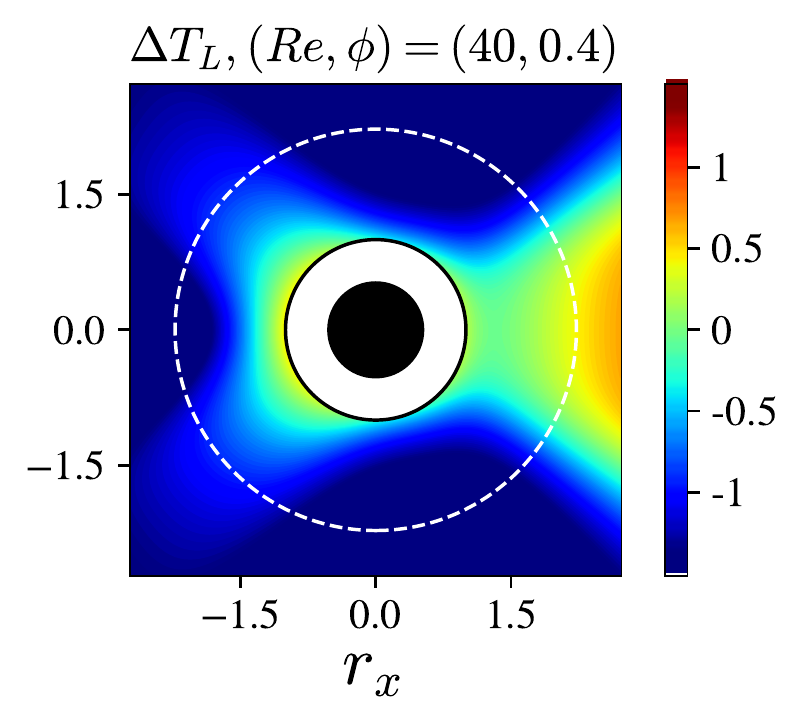}
		\label{fig:NN_out_dT_L}
	\end{subfigure}
	\caption{Scalar output of the NN blocks show in \cref{fig:PINN} after training of the PINN force and torque model at $ \Re = 40 $ and various $ \phi $. Note that these plots are obtained by taking the azimuthal average of of the NN block outputs over several planes that contain the streamwise direction vector $ \hat{\be}_x $. The black disk shows the location of a test particle, whereas the white annular band asserts that particles cannot overlap. The white dashed circles in each plot show the distance of the farthest neighbor to the reference particle.}
	\label{fig:NN_out}
\end{figure}
In order to demonstrate the interpretability of the PINN model, we show in \cref{fig:NN_out} the output of the NN blocks after training the model for predicting drag, lift and torque at $ \Re = 40 $ and various solid volume fractions. This is done by disconnecting the model at the summation junction in \cref{fig:PINN}, and extracting the outputs for arbitrary sets of $ r_x $, $ r_y $ and $ r_z $ values. Note that each NN block takes as input the $ x $, $ y $, and $ z $ relative coordinates of a neighbor $ j $ and outputs a scalar value. Since the visualization of contour values of a function in three dimensions is not feasible, we have computed the values of the function on several planes that contain $ \hat{\be}_x $ and averaged the resulting values in the azimuthal direction. This results in two-dimensional contour plots shown in \cref{fig:NN_out}. It is also for this reason that the distances in the transverse direction are shown as $ r_L $ rather than $ r_y $ or $ r_z $. We should point out that the output of the NN blocks on each plane is not fully axisymmetric, even though the contours shown in \cref{fig:NN_out} appear to be so owing to the averaging operation. This is due to the fact that there is no axisymmetry-enforcing mechanism in the model, neither is the dataset dense enough for the NN to learn the expected axisymmetry. This matter will be discussed later in this section.

The patterns observed in \cref{fig:NN_out} are reminiscent of the force maps of \cite{Moore2019}, influence maps of \cite{Balachandar2020}, as well as probability distribution maps of \cite{SeyedAhmadi2020}, indicating that the PINN model has successfully learned the physical patterns in the dataset. In \cref{fig:NN_out_dF_x}, the extended wake effect in the streamwise direction is prominently seen when the PINN is trained to predict drag. If a neighbor is located upstream of a test particle, the shielding effect substantially diminishes the drag experienced by the test particle, even when the separation distance between the two particles is large. Similarly, the increased pressure region in front of a downstream neighbor acts to decrease the fore-aft pressure difference of the test particle \cite{Akiki_2016}, hence decreasing its drag. Moreover, we have shown in \cite{SeyedAhmadi2020} that increased drag is associated with the lateral positioning of the neighboring particles, which can also be clearly identified in the positively valued areas located laterally with respect to the test particle in \cref{fig:NN_out_dF_x}.
In \cref{fig:NN_out_dF_L}, it can be seen that neighbors positioned laterally and downstream of the test particle induce a positive contribution to lift, whereas those located laterally and upstream create weaker negative contribution. It is important to note that the actual vectorial contribution of each neighbor is obtained by multiplying the scalar values given by the NN blocks by its corresponding influence vector, $ \hat{\be}_{L, j} $ in this case. This means that even though the upper and lower lateral downstream areas of force influence in \cref{fig:NN_out_dF_L} are both positive, a neighbor located in the region with $ r_L > 0 $ induces a lift force in the negative $ r_L $ direction, and vice versa. It is also worth mentioning that a neighbor $ j $ located downstream of the test particle with $ r_L = 0 $ would induce a zero lift influence, since $ \hat{\be}_x \times \hat{\be}_{j} = \bm{0} $ and hence $ \hat{\be}_{L, j} = \bm{0} $.
Finally, the NN block output for the torque in \cref{fig:NN_out_dT_L} shows positively valued regions immediately upstream and downstream of the test particle, accompanied by strong negatively valued areas that are positioned laterally. This pattern is in general qualitative accordance with our results in \cite{SeyedAhmadi2020}, although the strong negative areas are specific to the highest solid volume fraction of $ \phi = 0.4 $ shown here. The positively valued area downstream of the test particle is out of the range of neighbor distances that the model is trained for, and is thus an artifact.

A point worth further discussion is the explicit enforcing of different types of symmetry present in the physics of the problem. We mentioned earlier that the functional forms learned by the PINN model are not automatically axisymmetric; a condition that arises from the pairwise interactions assumption. One way to achieve symmetry and rotational invariance is data augmentation, which has been utilized in ML models of turbulence \cite{Ling2016a,Wu2018}. In order to achieve rotational symmetry along an arbitrary axis (i.e. axisymmetry), the training dataset can be rotated a certain number of times about the desired axis, and the resulting transformed data are added to the original dataset. In this manner, the ML algorithm encounters the same data in various transformed coordinates and is therefore better guided to respect the rotational symmetry in the problem. With our PINN model, the data augmentation technique with $ 10 $ to $ 30 $ rotations about the flow direction does result in improved---yet far from perfect--- axisymmetry of the NN block outputs, as well as a marginal performance enhancement. Nevertheless, the computation costs grow significantly with more rotations and hence prohibited further experimentation. Another conceivable method for enforcing axisymmetry would be to use the automatic differentiation technique \cite{Baydin2017} in order to mathematically impose the axisymmetry condition, i.e. to explicitly enforce the vanishing of the azimuthal gradient of the resulting NN block functions. This strategy is not implemented here, but seems to be a promising alternative for exploration in future works.

\begin{figure}[t]
	\centering
	\begin{subfigure}[t]{0.32\textwidth}
		\caption{}
		\vspace{2pt}
		\includegraphics[width=0.94\textwidth]{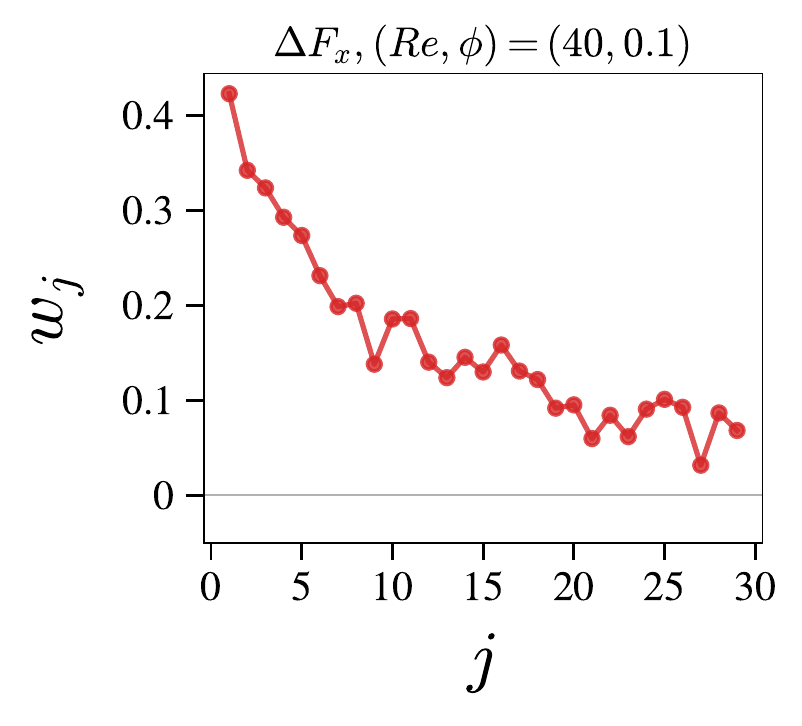}
		\label{fig:w_j_dF_x}
	\end{subfigure}
	\quad
	\begin{subfigure}[t]{0.3\textwidth}
		\caption{}
		\vspace{2pt}
		\includegraphics[width=0.9\textwidth]{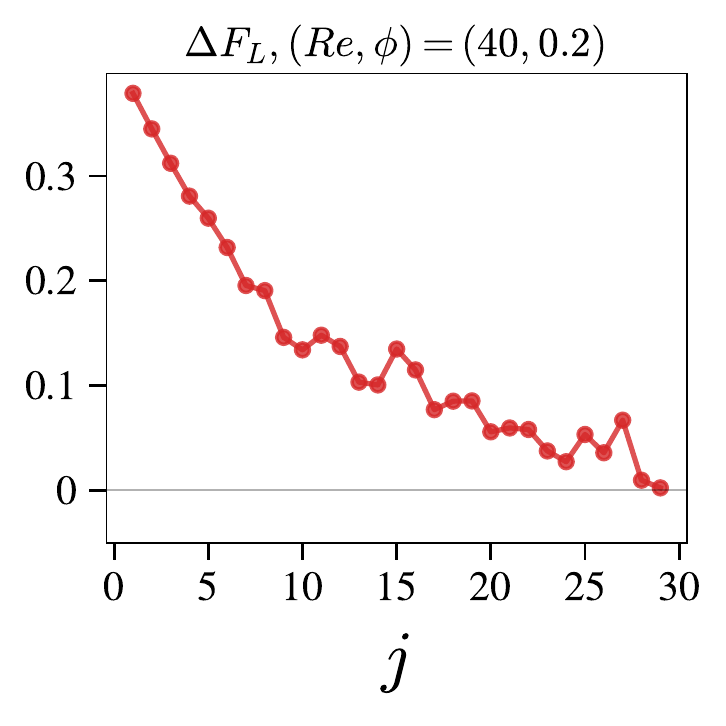}
		\label{fig:w_j_dF_L}
	\end{subfigure}
	\begin{subfigure}[t]{0.3\textwidth}
		\caption{}
		\vspace{2pt}
		\includegraphics[width=0.9\textwidth]{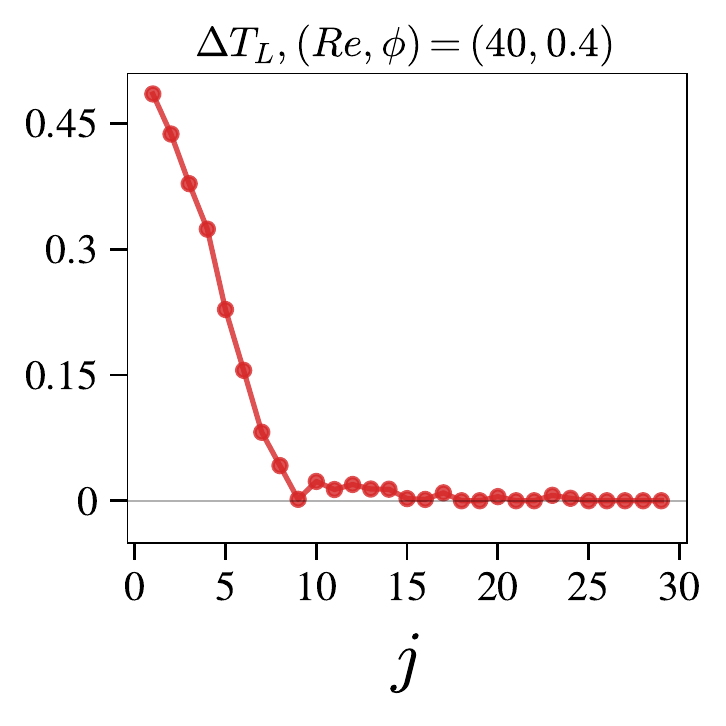}
		\label{fig:w_j_dT_L}
	\end{subfigure}
	\caption{Weighting parameters pertaining to the output layer of the PINN, after the training process for drag, lift and lateral torque. The Reynolds number is fixed at $ \Re = 40 $, whereas the solid volume fraction is varied from $ 0.1 $ to $ 0.4 $. The horizontal axis, $ j $, identifies the number of the neighboring particle.}
	\label{fig:w_j}
\end{figure}

More insight into the PINN model's interpretability can also be gained through the inspection of the weights in the output layer connections in \cref{fig:PINN}. Since the influence of each neighboring particle is separately handled by the model, the weight parameters in the last layer are each dedicated to the influence of a single neighbor. Given the linear activation function of the output layer, the final predicted value of the PINN is the weighted sum of individual neighbor influences.  Therefore, the magnitude of each weight parameter, denoted by $ w_j $, of the neighbor $ j $ reflects the importance of its contribution to the total force or torque deviation. In \cref{fig:w_j}, we show the values of weights in the output layer after training the PINN for force and torque prediction at $ \Re = 40 $ and various $ \phi $. In all cases, the decreasing trend of $ w_j $s with $ j $ is a clear evidence of the fact that neighbors located farther away have less influence on the force and torque deviation experienced by the test particle. Furthermore, $ w_j $ vanishes almost completely for $ j \geqslant 25 $ for this particular flow regime, whereas $ w_{j \geqslant 8} \approx 0 $ in case of the lateral torque indicating a much smaller radius of influence.
These results corroborate the observations made in \cref{fig:Rsq_M}; adding more neighbors located farther way to the model is unlikely to significantly enhance the model performance due to their diminishing influence. Note that in \cref{fig:w_j} the value of $ M $ (i.e. the number of included neighbors) is fixed, whereas $ M $ is variable in \cref{fig:Rsq_M} and a different set of $ w_j $ results from training with a particular value of $ M $.
\section{Concluding remarks}\label{sec:conclusion}
It is without doubt that computational methods have played a central role in shaping our understanding of multiphase flows. Today, the most complex scenarios of particle-laden flows involving the interaction of thousands of particles with each other and with the carrier fluid are accurately simulated using the PR-DNS technology. Nonetheless, PR-DNS is computationally intensive, and large-scale modeling of dense fluid-particle systems occurring in industrial or natural settings using PR-DNS is out of reach for the foreseeable future. The Euler-Lagrange technique provides a viable alternative by averaging the governing equations of the fluid phase and treating particles as point sources and sinks of momentum, leading to substantial reduction of the computational costs. The averaged equations, nevertheless, need to be supplemented with an appropriate closure law that accounts for the momentum exchange between the fluid and solid particles. Classical drag correlations proposed in the past based on PR-DNS of stationary arrays of particles only provide the average drag as a function of the Reynolds number and solid volume fraction. However, the force acting on each individual particle is also a function of the unique neighborhood of each particle. Furthermore, the average nature of the conventional closure laws precludes the computation of microstructure-induced lift and torque. A number of important efforts have been recently made to address this problem via the development of physics- and data-driven microstructure-informed models \cite{Akiki_2017,Moore2019,SeyedAhmadi2020,Balachandar2020}.

This paper presents a novel approach towards the direct modeling of hydrodynamic forces and torques in stationary beds of spheres using ML techniques. The tremendous success of ML algorithms in general pattern recognition tasks in the past few years has attracted significant interest in computational fluid dynamics research for the application of these methods to challenging problems in the field, including turbulence and reduced-order flow modeling.
The present work was motivated by the findings of \cite{Balachandar2020} that a conventional multilayer perceptron (i.e. fully connected NN) cannot be trained to give accurate predictions of the forces experienced by individual particles in a fixed bed of spheres. We remedy this issue by developing a NN model that incorporates---and essentially imitates---the physical formulation of the problem in its architecture, unlike a typical FCNN that is unaware of the physical nature of the problem at hand. Our physics-inspired NN model achieves substantial performance improvement over a conventional FCNN, while also enjoying full interpretability.
To this end, we invoke the pairwise interactions assumption that considers the influence of each neighbor on the force or torque of a test particle separately. The individual neighbor influences are ultimately superposed in order to obtain the total force or torque deviation from the mean value. Of particular importance is that this assumption pre-determines the direction of each influence vector. We embed the pairwise interactions assumption in the structure of the PINN by dedicating a NN block of fully connected layers that receives the position vector of a single neighbor as input and outputs a scalar value, multiplication of which by the already known influence vector yields the contribution of that particular neighbor. In this manner, the NN block is forced to learn the functional form of the neighbor influences according to the structure and constraints imposed on the model. The total force or torque deviation is then obtained by computing the weighted sum of the individual influences of each neighbor.
In addition to the pairwise interactions assumption, we also reduce the model complexity by sharing parameters between the NN blocks, which results in a unified functional representation of the force and torque influences. This is essentially a strategy similar to that employed in CNNs to manage complexity and enhance the generalizability of NN models. While handling inputs (i.e. position vectors) associated with different neighbors separately, parameter sharing results in the training of one, rather than several, unified functional form for the dependence of influences on the neighbor position.

The results show that the PINN model's predictions are on average as accurate as those of the MPP \cite{SeyedAhmadi2020} and PIEP \cite{Akiki_2017}, both of which similarly assume binary hydrodynamic interactions between particles, and hence only account for first-order effects of the local microstructure. We record the coefficient of determination $ \Rsq $ of the PINN model ranging from $ 0.51 $ to $ 0.89 $, with averages of $ 0.68 $, $ 0.64 $, and $ 0.73 $ for the drag, lift and lateral torque, respectively. The PINN model also proves to generalize well with the effective prevention of overfitting: the model does not grow in complexity with the inclusion of more neighbors in the training process, as directly opposed to the FCNN model.
In line with \cite{Balachandar2020}, we also demonstrate that the FCNN model fails to provide accurate predictions and inevitably suffers from overfitting when more than $ 3-4 $ neighbors are used in the training process. The worst case scenario pertains to the case of drag prediction where a maximum $ \Rsq $ of $ 0.28 $ was achieved with the FCNN model, whereas higher values of $ \Rsq $ reaching $ 0.39 $ for lift and $ 0.52 $ for lateral torque were recorded in selected cases. We attribute the better accuracy of the FCNN in the latter cases to the fact that the lift and torque of a test particle are less prone to long-range interactions compared to drag. In other words, their variations are explained by taking a smaller number of neighbors into account, thus yielding better predictions before the FCNN model becomes overfitted.
In addition to enforcing governing laws and constraints, an important hallmark of the effort to make ML algorithms aware of the underlying problem physics is to make such models interpretable, as contrasted with the black-box nature of a conventional multilayer perceptron. In the proposed PINN model, we have essentially achieved interpretability down to the level of the NN blocks that represent the functional form of neighbor influences. We have presented the intermediate outputs of the model generated by the NN blocks and demonstrated their correspondence to the governing fluid-particle physics, along with potential strategies for improvement. Finally, we also showed how the weighting parameters of the connections in the output layer reflect the relative importance of each neighbor's influence on the force and torque variation of a test particle.

In conclusion, we recognize that this model can only be as accurate as the assumptions based on which it is constructed. While being able to explain approximately two thirds of the total force and torque variation is considered a substantial improvement over classical closure laws, the performance reports of the present work and previous microstructure-informed models in \cite{Akiki_2017,Akiki2017,Moore2019,SeyedAhmadi2020,Balachandar2020} hint that we might have reached the maximum accuracy that can be achieved with models based on pairwise interactions. Further enhancement in model accuracy is only conceivable if second- and possibly higher-order interactions are accounted for in more sophisticated models.
Furthermore, another limiting factor is that current models are obtained based on data from simulations of stationary beds of spheres. In a time-dependent suspension of moving particles the interactions become even more complicated, as linear and angular velocities and accelerations of neighbors would also each play a role. Nevertheless, in such a scenario a PR-DNS can be run for extended periods of time in order to aggregate thousands of snapshots, which can then serve to significantly augment the dataset for training more complex ML models. While several possibilities are yet to be explored in future research on force and torque modeling, recent developments in the field appear to be very promising in high-fidelity up-scaling of PR-DNS using EL techniques.
\section*{Acknowledgments}
We greatly appreciate the financial support of the Natural Sciences and Engineering Research Council of Canada (NSERC) via their Discovery Grant Program. This research was enabled by support provided by Compute Canada (\url{http://www.computecanada.ca}) through Anthony Wachs's 2020 Computing Resources for Research Groups allocation qpf-764-ac.
\renewcommand*{\bibfont}{\footnotesize}
\printbibliography[heading=bibintoc]
\end{document}